\documentclass{emulateapj}
\usepackage{multirow}
\usepackage{graphicx}

\slugcomment{\it To be submitted to the Astrophysical Journal}
\shortauthors{Lin Yan et al.}
\shorttitle{Detections of CO Molecular Gas in Spitzer $z \sim 2$ ULIRGs}

\def\deg{\ifmmode {^{\circ}}\else {$^\circ$}\fi}
\def\kms{\ifmmode {\rm\,km\,s^{-1}}\else
    ${\rm\,km\,s^{-1}}$\fi}
\def\ergcm2s{\ifmmode {\rm\,ergs\,cm^{-2}\,s^{-1}}\else
    ${\rm\,ergs\,cm^{-2}\,s^{-1}}$\fi}
\def\ergAcm2s{\ifmmode {\rm\,ergs\,cm^{-2}\,s^{-1}\,\AA^{-1}}\else
    ${\rm\,ergs\,cm^{-2}\,s^{-1}\,\AA^{-1}}$\fi}
\def\ergs{\ifmmode {\rm\,ergs\,s^{-1}}\else
    ${\rm\,ergs\,s^{-1}}$\fi}
\def\kmsMpc{\ifmmode {\rm\,km\,s^{-1}\,Mpc^{-1}}\else
    ${\rm\,km\,s^{-1}\,Mpc^{-1}}$\fi}

\def\spose#1{\hbox to 0pt{#1\hss}}
\def\simlt{\mathrel{\spose{\lower 3pt\hbox{$\mathchar"218$}}
     \raise 2.0pt\hbox{$\mathchar"13C$}}}
\def\simgt{\mathrel{\spose{\lower 3pt\hbox{$\mathchar"218$}}
     \raise 2.0pt\hbox{$\mathchar"13E$}}}

\def\gs{\mathrel{\raise0.35ex\hbox{$\scriptstyle >$}\kern-0.6em
\lower0.40ex\hbox{{$\scriptstyle \sim$}}}}
\def\ls{\mathrel{\raise0.35ex\hbox{$\scriptstyle <$}\kern-0.6em
\lower0.40ex\hbox{{$\scriptstyle \sim$}}}}

\newcommand{\um}{\,$\mu$m}

\newcommand{\spitz}{{\sl Spitzer }}

\begin{document}

\title{Detections of CO Molecular Gas in 24\,micron-Bright ULIRGs at $\sl{z}$\,$\sim$\,2 in the \spitz First Look Survey\footnote{Based on observations obtained at the {\it Institute for Radioastronomy at Millimeter Wavelengths } (IRAM) Plateau de Bure Interferometer (PdBI). IRAM is funded by the Centre National de la Recherche« Scientifique (France), the Max-Planck Gesellschaft (Germany), and the Instituto Geografico Nacional (Spain).}}

\author{Lin Yan,$^2$ L.~J. Tacconi,$^3$, N. Fiolet,$^4$ A. Sajina,$^5$ A. Omont,$^4$ D. Lutz,$^3$ M. Zamojski,$^2$ R. Neri,$^6$ P. Cox,$^6$ K.~M. Dasyra,$^2$}

\affil{$^2$ \spitz Science Center, California Institute of Technology, MS 220-6, Pasadena, CA 91125, USA}
\affil{$^3$ Max-Planck Institut f\"{u}r Extraterrestrische Physik (MPE),Giessenbachstrasse, D-85741 Garching, Germany}
\affil{$^4$ UPMC University of Paris 06, UMR7095, Institut d'Astrophysique de Paris, F-75014, Paris, France}
\affil{$^5$ Haverford College, Haverford, PA 19041, USA}
\affil{$^6$ Institut de Radio Astronomie Millimetrique (IRAM), St. Martin dÕHeres, France}
\affil{$^7$ NASA Herschel Science Center, California Institute of Technology, MS 220-6, Pasadena, CA 91125, USA}
\email{lyan@ipac.caltech.edu}

\begin{abstract}

We present CO observations of  nine ULIRGs at $z$\,$\sim$\,2 with $f_\nu(24\mu m)$\,$\simgt$\,1\,mJy,  previously confirmed with the mid-IR spectra in the {\sl Spitzer} First Look Survey.  All targets are required to have accurate redshifts from Keck/GEMINI near-IR spectra. 
Using the Plateau de Bure millimeter-wave Interferometer (PdBI) at IRAM,  we detect CO J(3-2) [7 objects] or J(2-1) [1 object] line emission from eight sources with integrated intensities $I_{c}$\,$\sim$\,(5\,--\,9)$\sigma$.  
The CO detected sources have a variety of mid-IR spectra, including strong PAH, deep silicate absorption and power-law continuum, implying that these molecular gas rich objects at $z$\,$\sim$\,2 could be either starbursts or dust obscured AGNs.  The measured line luminosity $\rm L^{'}_{CO(3-2)}$ is (1.28\,--\,3.77)$\times$10$^{10}$\,K\,km/s\,pc$^2$.  The averaged molecular gas mass $\rm M_{H_2}$ is $1.7\times$10$^{10}M_\odot$, assuming CO-to-H$_2$ conversion factor of 0.8\,$M_\odot$/[K\,km/s\,pc$^2$]. 
Three sources (33\%) -- MIPS506, MIPS16144 \&\ MIPS8342 -- have double peak velocity profiles. The CO double peaks in MIPS506 and MIPS16144 show spatial separations of 45\,kpc and 10.9\,kpc, allowing the estimates of the dynamical masses of 
3.2$\times$10$^{11}$\,$\rm{sin}^{-2}(i)\,M_\odot$ and 5.4$\times$10$^{11}$\,$\rm{sin}^{-2}(i)\,M_\odot$ respectively. The implied gas fraction, $\rm M_{gas}/M_{dyn}$, is 3\%\ and 4\%, assuming an average inclination angle.
Finally,  the analysis of the HST/NICMOS images, mid-IR spectra and IR SED revealed that most of our sources are mergers, containing dust obscured AGNs dominating the luminosities at (3\,--\,6)\um.  Together, these results provide some evidence 
suggesting SMGs, bright 24\um\ ULIRGs and QSOs could represent three different stages of a single evolutionary sequence, however, a complete physical model would require much more data, especially high spatial resolution spectroscopy. 
\end{abstract}


\section{Introduction \label{sec:intro}}

Molecular gas holds one of the keys to fundamental questions about the formation and evolution of galaxies.  Particularly, at high redshift, significant galaxy growth in its stellar population and black hole occurs during dusty, gas rich phases.  The importance of infrared (IR) luminous, gas rich populations has been high-lighted by observations from the {\it Infrared Space Observatory} ({\it ISO}),  ground-based (sub)millimeter  cameras, and more recently, the {\it Spitzer Space Telescope} \citep{genzel00,blain02,lagache05,tom08}.  Particularly, \spitz dramatically improved our ability to probe high redshift ($z$\,$\simgt$\,1), dusty galaxies down to $L_{IR}$\,$\simgt$\,10$^{11}L_\odot$, increasing the number of detections of such sources by orders of magnitude in comparison with $\sim$\,(200\,--\,300) $z$\,$\sim$\,2 gas rich, sub-millimeter galaxies. 
Several recent studies using the \spitz InfraRed Spectrograph \citep[IRS;][]{houck04} have confirmed with mid-IR spectra a population of 24\um-selected, Ultra-luminous infrared galaxies (ULIRGs) at $z$\,$\sim$\,2 \citep{yan05,houck05,weedman06,yan07,sajina08,dey08,dasyra09}.  One important feature of these studies is that the samples are selected to have 24\um\ fluxes brighter than $\sim$\,1mJy,  placing them among the most IR luminous objects at $z$\,$\sim$\,2 with $L_{IR}$\,$ \sim$\,10$^{12.5-13.3} L_\odot$ \citep{sajina07,sajina08}.  Other gas rich galaxies with comparable or more $\rm L_{IR}$ at $z$\,$\simgt$\,2 are sub-millimeter galaxies (SMGs) and optically bright QSOs with CO detections.  By definition, SMGs are selected by their far-infrared emission, produced by larger, colder dust grains, in contrast, \spitz bright 24\um\ ULIRGs are selected in the mid-IR, probing smaller and hotter dust grains.

\begin{deluxetable}{lcccc}
\singlespace
\tablecolumns{5} 
\tablewidth{0in}
\tablecaption{Sample Targets \label{targets}}
\tabletypesize{\footnotesize}
\tablehead{
\colhead{ID} &
\colhead{RA} &
\colhead{DEC} & 
\colhead{z(Opt)$^a$} &
\colhead{z(CO)} \\
  & deg & deg &  &  
}
\startdata
MIPS429    &  259.04930   & 59.20366   &  2.213  & 2.2010   \\
MIPS506    &  257.91080   & 58.64405   &  2.470  & 2.4704   \\
MIPS8196   &  258.79282   & 60.16533   &  2.586  & ....$^b$     \\
MIPS8327   &  258.89908   & 60.47375   &  2.441  & 2.4421    \\
MIPS8342   &  258.54813   & 60.18591   &  1.562  & 1.5619  \\
MIPS15949  &  260.28842   & 60.25035   &  2.122  & 2.1194  \\
MIPS16059  &  261.11850   & 60.25922   &  2.326  & 2.3256  \\
MIPS16080  &  259.68655   & 60.02107   &  2.007  & 2.0063  \\
MIPS16144  &  261.09207   & 59.53077   &  2.131  & 2.1280     \\

\enddata
\tablenotetext{a}{The optical spectroscopic redshifts  are taken from \citet{sajina08}.}
\tablenotetext{b}{MIPS8196 has no significant CO line detection, thus, its CO redshift is not determined.}
\end{deluxetable}

SMGs, \spitz bright 24\um\ ULIRGs and gas rich QSOs are all high-$z$ systems with extremely high bolometric luminosities, leading 
to the important question of how these systems are related or different.  Studies of the gas content, particularly CO gas, of SMGs have made significant progress in recent years \citep{neri03, greve05, tacconi06, tacconi08}.  The general consensus for this population is that SMGs contain a reservoir of $10^{10-11} M_\odot$ of molecular gas, distributed over a small area of  $R_{1/2}$\,$\simlt$\,2\,kpc. This large amount of molecular gas is being converted to stars at a rate of $\sim1000M_\odot$/yr, typically over several dynamical time scales of  $\sim$\,$10^8$yrs.  Recent several studies have also examined the mid-IR spectral properties of SMGs \citep{lutz05b,valiante07,pope08,karin07,karin09}.  The one clear difference is that the observed 24\um\ fluxes of SMG  are on average  several 100\,$\mu$Jy, or a factor of (2\,--\,3) smaller than  their \spitz-selected counterparts.  This is consistent with the conclusion that bright 24\um\ $z$\,$\sim$\,2 ULIRGs are starburst/AGN composite systems, with more AGN heating the hot dust grains, thus generating elevated continuum emission at (3\,--\,6)\um\ than that of SMGs \citep{sajina08,polletta08}.  In contrast, for the majority of SMGs,  starburst dominates the IR luminosity, while AGN contributes only a small fraction of $\rm L_{IR}$  \citep{valiante07,pope08,dave08,karin09}. In addition,  abundant CO molecular gas has been detected among optically selected, high-$z$ QSOs and radio galaxies, especially gravitationally lensed systems. \citet{Solomon05} reviewed and compiled a list of 23 QSOs and radio galaxies which have at least low resolution CO line observations (also \citet{greve05} for a similar list).   


In this paper, we will present the measurement of cold molecular gas masses among bright 24\um\ ULIRGs at $z$\,$\sim$\,2, determine their gas dynamics, and understand how they differ from SMGs and gas rich QSOs \&\ radio galaxies.  To directly address these questions, we obtained CO interferometric observations for a sample of bright 24\um\ ULIRGs at $z$\,$\sim$\,2 selected from our previously published studies \citep{yan07, sajina08}.  This paper is organized as follows. \S\,2 describes the sample selection, ancillary data used for the analyses in this paper, and the CO interferometric observations and data reduction. \S\,3 presents the results, and discusses the physical implications by combining all available imaging and spectroscopic data. To understand how 24\um\ ULIRGs differs from SMGs and high-$z$ QSOs, we make comparisons to the QSOs and radio galaxy sample compiled by \citet{Solomon05} and also to a small subset of QSOs which have reliable dynamic masses from high resolution CO observations and robust black hole mass estimates from UV spectroscopy.  \S\,4 summarizes the main conclusions from this paper.  Throughout the paper, we adopt an $\Omega_M$\,=\,0.27, $\Omega_\Lambda$\,=\,0.73\ and $H_0$\,=\,71\kmsMpc\ cosmology \citep{spergel03}.

\section{Observational Data \label{data}} 

\begin{deluxetable*}{lcccccccccc}
\singlespace
\tablecolumns{11} 
\tablewidth{0in}
\tablecaption{CO observations \label{obs}}
\tabletypesize{\footnotesize}
\tablehead{
\colhead{ID} &
\colhead{Lines} &
\colhead{$\nu_{obs}$} &
\colhead{Config.} &
\colhead{Beam$^a$}  &
\colhead{PA} &
\colhead{Time$^b$} &
\colhead{Noise/chan.} & 
\colhead{Chan.width} &
\colhead{$\Delta^c$} &
\colhead{RMS$^d$} \\
  &  &  GHz &  &  &  & hours  & mJy/chan & km/s  & km/s & mJy/beam \\
}
\startdata
MIPS429  &  (3-2)  &  107.642  &   C  & $4.8^{''}\times3.89^{''}$  & $68^\deg$ &  6   &  0.4  & 83  & 664 & 0.2 \\
MIPS506  &  (3-2)  &  99.679   &   D  & $5.02^{''}\times4.37^{''}$ & $71^\deg$ &  4   &  0.35 & 75  & 525 & 0.25 \\
MIPS8196  &  (3-2)  &  96.456  &   C  & $5.6^{''}\times4.18^{''}$  & $65^\deg$ &  4.5 &  0.39 & 62  & ... & .. \\
MIPS8327  &  (3-2)  &  100.464 &   C  & $2.86^{''}\times2.5^{''}$  & $71^\deg$ &  9   &  0.7  & 60  & 540 & 0.12 \\
MIPS8342  &  (2-1)  &  89.913  &   D  & $6.35^{''}\times4.15^{''}$ & $56^\deg$ &  9.5 &  0.6  & 67  & 670 & 0.1 \\
MIPS15949  &  (3-2)  &  111.046  & C  & $2.63^{''}\times2.3^{''}$  & $69^\deg$ &  10.1 & 0.4  & 54  & 756 & 0.12 \\
MIPS16059  &  (3-2)  &  103.999  & C  & $4.77^{''}\times3.83^{''}$ & $89^\deg$ &  5.5  & 0.5  & 87  & 522 & 0.25 \\
MIPS16080  &  (3-2)  &  114.997  & C  & $4.61^{''}\times3.52^{''}$ & $75^\deg$ &  10   & 0.7  & 72  & 360 & 0.35 \\
MIPS16144  &  (3-2)  &  110.407  & C  & $2.44^{''}\times2.15^{''}$ & $62^\deg$ &  5.4  & 0.5  & 81  & 810 & 0.13 \\

\enddata
\tablenotetext{a}{Beam is defined as the semi-major axis times the semi-minor axis in arcseconds.}
\tablenotetext{b}{This is the total on-target integration time, calculated equivalent to 6 antenna.}
\tablenotetext{c}{The summed CO maps are integrated over a velocity width $\Delta $ with the center on the peak of the CO emission line.}
\tablenotetext{d}{This noise is RMS for the summed CO map made with the listed parameters.}
\end{deluxetable*}

The ground transition temperature for CO J(1-0) is only 5.5\,K, in contrast for H$_2$ 500\,K. Therefore, CO rotational emission lines are easily produced by collisional excitation.  It has been shown that  CO flux  $I_{CO}$ linearly correlates with the column density of molecular hydrogen H$_2$, providing a sensitive tracer of  bulk of the cold molecular gas in the Universe \citep{young82,dickman86,solomon87}.  Besides the molecular gas content of a galaxy, CO interferometric observations can also probe the spatial distribution as well as velocity field of molecular gas. Therefore, we obtained 
new CO emission line observations for nine bright 24\um\ ULIRGs at $z$\,$\sim$\,2 using the PdBI at IRAM.  In this section, we describe the selection of our CO targets, and summarize the CO observations and all of the ancillary all of the data. 

\subsection{The sample and ancillary data}

The nine targets observed by the PdBI were selected from a large sample of  $z$\,$\sim$\,1\,--\,2 ULIRGs with \spitz mid-IR spectra published in \citet{yan07}.  This parent sample consists of 52 sources initially selected with 24\um\ flux density brighter than 0.9\,mJy, and very red 24-to-8\um\ and 24-to-R colors from the 4\,square degree \spitz Extragalactic First Look Survey (XFLS)\footnote{$R(24,8)$\,$\equiv$\,$\log_{10}(\nu f_{\nu}(24\mu m)/\nu f_{\nu}(8\mu m)$\,$\simgt$\,0.5; \\ $R(24,0.7)$\,$\equiv$\,$\log_{10}(\nu f_{\nu}(24\mu m)/\nu f_{\nu}(0.7\mu m)$\,$\simgt$\,1.0}.  The subsequently obtained \spitz mid-IR spectra covering the rest-frame 4\,--\,20\um\ determined that 74\%\ of the sample is at 1.5\,$<$\,$z$\,$<$\,3.2, confirming the effectiveness of the initial color selection \citep{yan04}.  The total infrared luminosities ($\rm L_{3-1000\mu m}$) are in the range of $10^{12-13}L_\odot$.  The complete mid-IR spectral and SED analysis shows that at least $\sim$75\%\ of the sample contain dust obscured AGNs, which dominate the mid-IR (3\,--\,6)\um\ luminosities but star formation still contribute most of the far-IR luminosities \citep{sajina07, sajina08}.  Of the 52 sources, 44 galaxies were observed at 1.2\,mm with the IRAM 30\,meter telescope using the 117 element version of the MAMBO array \citep{kreysa98}.  Of these, 7 are detected at $\simgt3\sigma$ (14 at $\simgt2\sigma$), with an average $\rm{rms}\sim0.6$\,mJy.  The 1.2\,mm data of the full sample has been published in \citet{lutz05} and \citet{sajina08}.

To measure cold molecular gas content and to quantify how it changes with mid-IR properties among these $z$\,$\sim$\,2 \spitz ULIRGs, we obtained CO interferometric observations for sources with a broad range of mid-IR spectral properties, including sources with strong PAH emission, deep silicate absorption,  and mid-IR power-law continua.  The primary selection of our CO targets was the availability of accurate spectroscopic redshifts from Keck and GEMINI.  This criterion has limited our targets to  a small subset of the full sample.  All of our targets have 1.2\,mm observations from MAMBO.  Of the nine sources, only one have 1.2\,mm SNR\,$>$\,$3\sigma$, seven with SNR\,$\sim$1.81\,--\,2.58, and one with SNR\,$<$\,1.5.  Excluding the $>$\,$3\sigma$ source, we stacked the remaining 8 sources, yielding an averaged 1.2\,mm flux of 1.06\,$\pm$\,0.18\,mJy.  This value is higher than the averaged value of 0.5\,$\pm$\,0.1\,mJy for the 44 sources observed at 1.2\,mm in the original \spitz sample (52 objects) \citep{sajina08}.  This implies that our CO targets systematically have more far-IR emission, in comparison with the full \spitz mid-IR spectroscopic sample at $z$\,$\sim$\,2. 

Table~\ref{targets} and Table~\ref{obs} list the coordinates, redshifts (from both CO and near-IR spectra) and the CO observational parameters for the 9 sources observed by PdBI.   Table~\ref{ancillary} shows the ancillary observations from optical, near-IR, far-IR and 20\,cm photometry \citep[see][for details]{sajina08}. 
The near-IR spectra were obtained using NIRSPEC at the Keck and NIRI at the GEMINI.  We have high resolution H-band images from {\it HST} NICMOS.  This dataset has been published in \citet{dasyra08} and is included in the analysis of a larger sample of $z$\,$\sim$\,0.3\,--\,2 bright 24\um\ ULIRGs in \citep{michel09}.

\subsection{CO emission line observations}

The observations were carried out in 2007-2008 with the PdBI in D (2 objects) and C (7 objects) configuration in order to maximize the sensitivity.  The 3-mm receivers were tuned to the central observed frequency according to the Near-IR spectroscopic redshifts.  Of the total nine sources,  we used the PdBI to observe the CO J(3-2) transition ($\nu_{rest}$\,=\,345.796\,GHz) for eight objects and the J(2-1) transition ($\nu_{rest}$\,=\,230.538\,GHz) for one object at redshift of $1.562$.  Typically, our sources were observed under good weather conditions with 1\,--\,2 tracks for a total of (5\,--\,10)\,hours, 6 antenna equivalent, on-target integration.   Table~\ref{obs} lists the observation parameters. The data reduction takes places in two stages using the IRAM GILDAS software.  First, the raw data are calibrated using the software CLIC, developed at IRAM\footnote{A complete documentation is available at
http://www.iram.fr/IRAMFR/GILDAS/doc/html/clic-html/clic.html}.  Care is given to the phase, flux and amplitude calibrations by rejecting any calibration anomalies due to meteorological conditions or electronics.  After calibration, CLIC generates the visibility data, {\it i.e.}``uv'' tables. Second,  based on these uv tables, we extract CO maps and spectra using the IRAM MAPPING software\footnote{The documentation is at \\ 
http://www.iram.fr/IRAMFR/GILDAS/doc/html/map-html/map.html}.  The synthesized, clean beam size is typically elongated, roughly (2.4$^{''}$$\times$2.4$^{''})$ to (5.6$^{''}$$\times$4.18$^{''})$ for the C configuration and (5.02$^{''}$$\times$4.37$^{''})$ to (6.2$^{''}$$\times$4.15$^{''})$  for the D configuration.  The corresponding spatial resolution in linear size ranges from 20.5 to 53\,kpc. For comparison, for the \spitz 24\um\ band, the Full Width Half Maximum (FWHM) of a point source is $6^{''}$ and the NICMOS H-band PSF is $0.22^{''}$.   

\begin{figure*}[!t]
\centering{\includegraphics[width=2.2\columnwidth]{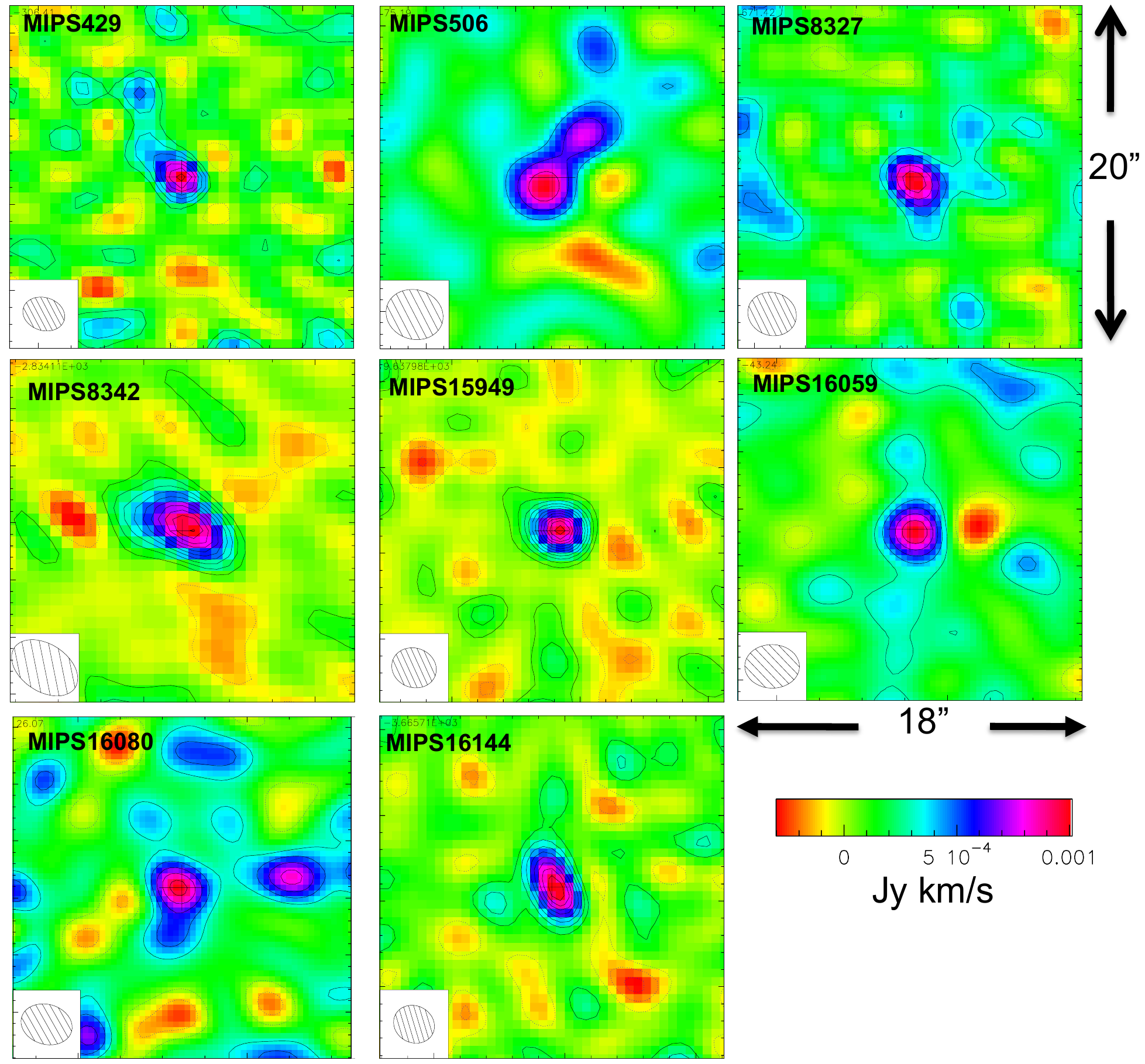}}
\caption{\footnotesize The integrated CO maps for the 8 detected sources. The small insert in each panel shows the size and shape of the beam. The cross marks the center of 24$\mu m$ source.  The first contour starts at 1.17$\sigma$, 1.15$\sigma$, 1.05$\sigma$, 1.05$\sigma$, 1.04$\sigma$, 1.18$\sigma$, 1$\sigma$, and 1.08$\sigma$ respectively for MIPS429, MIPS506, MIPS8327, MIPS8342, MIPS15949, MIPS16059, MIPS16080 and MIPS16144.  The contour step size is 1$\sigma$ for all maps. The rms $\sigma$ (mJy/beam), the channel width and the number of channels over which the CO maps are integrated, are given in Table~\ref{obs}. \label{comap}}
\end{figure*}

Figure~\ref{comap} shows the integrated CO maps, with the small insert at the left bottom of each panel indicating the beam size and positional angle.  Before producing the summed CO maps, we rebinned the data with an original 10\,MHz channel width by a factor of (3\,--\,4), yielding smoothed spectral data cubes with a channel width roughly (54\,--\,83)\,km/s.  The CO maps shown in Figure~\ref{comap} were made by  summing over a number of velocity channels, ranging from 5\,--\,14 (equivalently, velocity width $\Delta$\,$\sim$\,360\,--\,810\,km/s in Table~\ref{obs}), centered on the peak CO line emission. The final maps have $1\sigma$ noise values ranging from 0.1 to 0.35\,mJy/beam.  Table~\ref{obs} lists the noise parameters and the parameters used for making the summed CO maps.  In Figure~\ref{comap}, the first CO contour starts roughly at 1$\sigma$ with a step size of 1$\sigma$ for all panels.  The summed CO line emissions are detected at (4.7\,--\,9.7)$\sigma$ for 8 sources.  Table~\ref{coparam} lists the integrated CO line fluxes,  velocity width over which the CO line integration is done,  and derived CO luminosity and H$_2$(+He) molecular masses. 

The CO line luminosity can be derived from the line intensity, following Solomon et al. (1997):

\begin{equation}
L_{CO} [L_\odot] = 1.04 \times 10^{-3} \Big({\rm I_{CO} \over \rm Jy\,km\,s^{-1}}\Big) \Big({\nu_{obs} \over \rm GHz}\Big) \Big({\rm D_L \over \rm Mpc}\Big)^2
\end{equation}

and

\begin{equation}
L^{'}_{CO} [\rm K\,km/s\,pc^{2}] = 3.25 \times 10^7 {I_{CO}} {\nu_{obs}}^{-2} {D_L}^2 (1+z)^{-3},
\end{equation}

Here $L^{'}_{CO}$ is the CO line luminosity in units of K\,km/s\,pc$^{2}$, $I_{CO}$ is the integrated line intensity in Jy\,km/s,  $\rm D_L$ is the luminosity distance in units of Mpc, $\nu_{obs}$ is the observed frequency in GHz.  The H$_2$ (+He) molecular gas mass can be estimated using the following equation:

\begin{equation}
{M_{gas} \over M_\odot} = \alpha \Big({T_{CO(3-2)}/T_{CO(1-0)} \over 1}\Big) L^{'}_{CO}
\end{equation}
 
Here $\alpha$ is the CO J(1-0) luminosity to molecular gas mass conversion factor in units of $M_\odot \rm [K\,km/s\,pc^{2}]^{-1}$ and $T_{CO}$ is the effective brightness temperature of a CO transition.  Here we assume $T_{CO(3-2)}/T_{CO(1-0)} = 1$ and $\alpha$\,=\,$\rm 0.8\,M_\odot (K\, km/s\,pc^{2})^{-1}$ \citep{solomon97,downes98}, the same assumption used for SMGs and QSOs\citep{greve05,tacconi06,tacconi08}.  Although a more realistic $T_{CO(3-2)}/T_{CO(1-0)}$ ratio is probably between (0.5\,--\,0.7) \citep{wilson08}, we adopt our assumption in order to make direct comparison without additional corrections to the published SMG and QSO data.


\subsection{Modeling the IR SED \label{sec:sed}}

\begin{figure*}[htbp]
\centering{\includegraphics[width=2.\columnwidth]{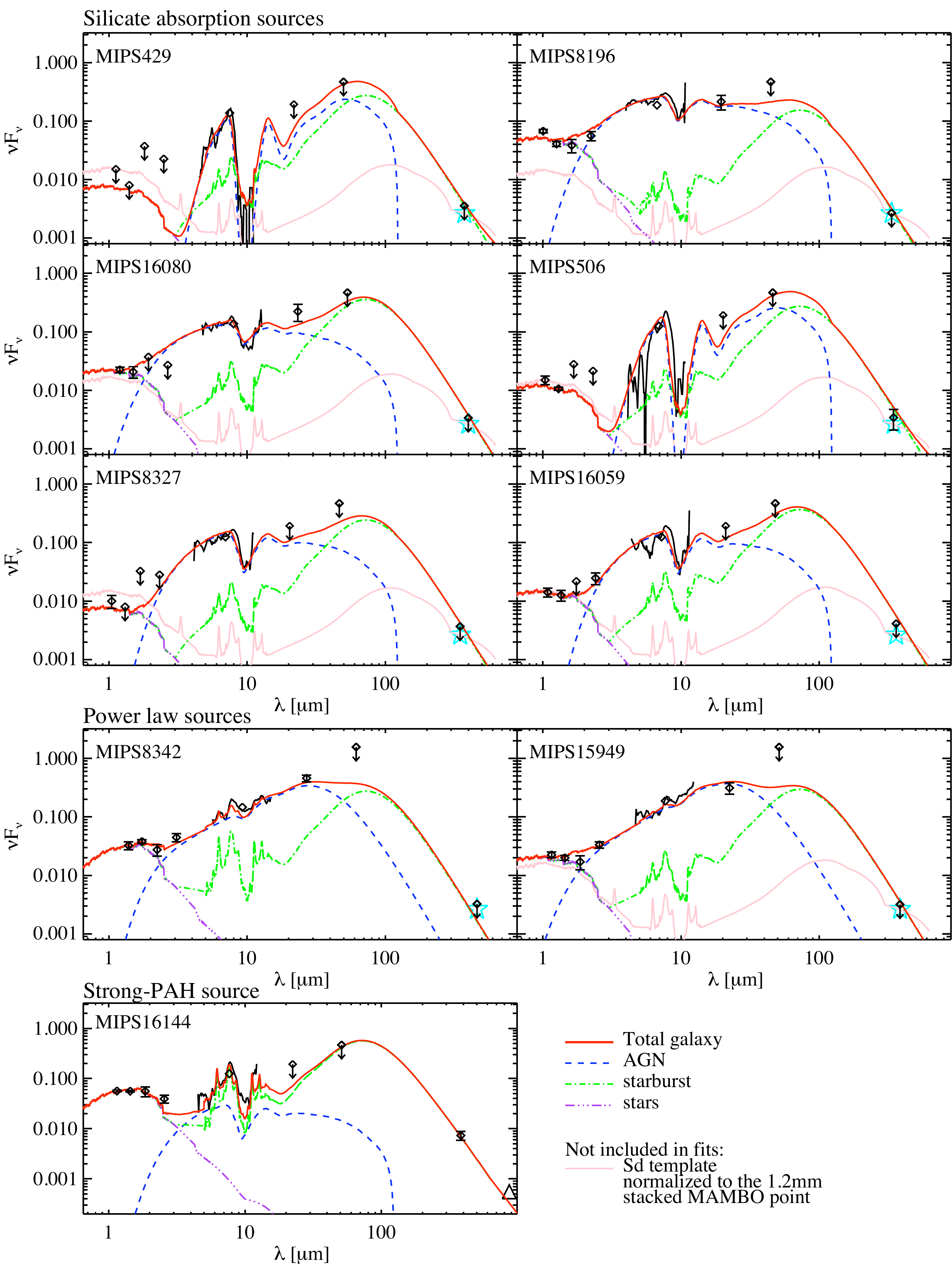}}
\caption{\footnotesize Near-to-far-IR spectral energy distributions for the 9 targets.  The figure legend shows the starburst, AGN and stellar components.  The large cyan stars at the observed 1.2\,mm mark the averaged flux value of 1.06\,$\pm$\,0.18\,mJy. The down-ward arrows mark the limits of undetected sources.  For MIPS16144, the open black triangle indicates the rest-frame continuum flux observed in the CO J(3-2) spectrum.\label{sed}}
\end{figure*}

To derive the total and far-IR luminosities ($\rm L_{IR}$\,=\,$\rm L_{3-1000\mu m}, L_{FIR}$\,=\,$\rm L_{40-120\mu m}$), we fit the spectral energy distribution of our sources from near-IR to far-IR (see Table~\ref{ancillary}).  The longest wavelength data available is MAMBO 1.2\,mm, which is crucial for constraining the far-IR emission.  All of the nine targets except one have weak 1.2\,mm fluxes at $3\sigma$ or less.  For these 8 sources, the sigma weighted, averaged flux is 1.06\,$\pm$\,0.18\,mJy (5.9$\sigma$), providing constraining power in the SED fitting in Figure~\ref{sed} (cyan, open stars).  
Our SED fitting uses three components:  pure starburst template (the averaged SMG SED from \citet{pope06,pope08}, green dot-dashed line), reddened type-I quasar SED (\citet{richards06}, blue dashed line), and a 2\,Gyr old Single Stellar Population (SSP) model from \citet{maraston05} to account for the near-IR emission (purple dot-dot-dashed line).  The solid red line is the sum of these three components.  The type I quasar SED (blue) has to be reddened substantially ($A_V \simgt 5$) assuming a MW-type extinction \citep{draine03}. For most sources this leads to a reasonably good fit to the mid-IR continuum and silicate absorption feature. However, two of our sources (MIPS8342 and MIPS15949) show red mid-IR continua but no significant silicate absorption.  The above approach cannot fit this type of spectra. For these sources, we revert to our power-law continuum approach as used in \citet{sajina08}.  
\begin{deluxetable*}{lccccccccc}
\singlespace
\tablecolumns{10} 
\tablewidth{0in}
\tablecaption{Ancillary data and derived parameters for the targets \label{ancillary}}
\tabletypesize{\footnotesize}
\tablehead{
\colhead{ID} &
\colhead{R} &
\colhead{H} & 
\colhead{Flux} & 
\colhead{Flux$^a$}  &
\colhead{Flux$^a$} &
\colhead{Flux}  &
\colhead{Flux}  &
\colhead{Flux}  \\
  & vega & AB & 24$\mu m$ & 70$\mu m$ & 160$\mu m$& 1.2mm  & 1.4GHz  &  650MHz \\
  & mag & mag & mJy             & mJy             & mJy             & mJy        &  mJy       &   mJy       \\ }
\startdata
MIPS429   & $<$25.5  &  ...  &  1.10 & $<$4.5 & $<$30  & 1.03$\pm$0.57    &  $<$0.08 & ... \\ 
MIPS506   & 23.44 & 21.66 &  1.06 & $<$4.5 & $<$30  & 1.37$\pm$0.53    &  $0.14\pm0.03$    & ... \\
MIPS8196  & 22.45 & 18.68 &  1.50 & $5.0\pm1.4$ & $<$30   & 0.99$\pm$0.43   &  $<$0.08 & ... \\
MIPS8327  & 23.27 & 21.06 &  1.16 & $<$5.8 & $<$30   & 1.03$\pm$0.59   &  $1.4\pm0.06$     & $3.44\pm0.14$ \\
MIPS8342  & 23.51 & 21.54 &  1.17 & $10.7\pm1.4$ & 29$\pm$11$^b$ & 0.98$\pm$0.52  & $0.18\pm0.03$   & $0.86\pm0.11$  \\
MIPS15949 & 22.90 & 20.73 &  1.50 & $7.3\pm$1.6 & $<$30   & 1.24$\pm$0.51   & $0.16\pm0.02$     & ...     \\
MIPS16059 & 23.48 & 20.49 &  1.29 & $<7.2$ & $<$30   & 1.20$\pm$0.66  & $0.57\pm0.03$     & ...     \\
MIPS16080 & 22.86 & 20.31 &  1.10 & $5.2\pm1.7$ & $<$30   & 0.69$\pm$0.54  & $0.34\pm0.03$     & ... \\
MIPS16144 & 23.10 & 20.85 &  1.12 & $<3.6$ & $<$30   & 2.93$\pm$0.59  & $0.12\pm0.03$     & ...    \\

\enddata
\tablenotetext{a}{For none-detections at 70\um\ and 160\um, the fluxes are listed as 3$\sigma$.}
\tablenotetext{b}{MIPS8342 70\um\ position corresponds to multiple 24\um\ sources. Here we deblended the 70\um\ blob, and list only the partial flux belonging to MIPS8342 with the flux error from the total. See Sajina et al. (2008) for detail.}
\end{deluxetable*}

\begin{deluxetable*}{lccllcc}
\singlespace
\tablecolumns{7} 
\tablewidth{0in}
\tablecaption{Derived CO parameters \label{coparam}}
\tabletypesize{\footnotesize}
\tablehead{
\colhead{ID$^a$} &
\colhead{FWHM1$^b$}  &
\colhead{FWHM2$^c$} &
\colhead{I$_{co}$} &
\colhead{L$_{co}$} &
\colhead{L$^{'}_{co}(10^{10})^d$} &
\colhead{Mass(H$_2$)$^d$} \\ 
    &  km/s   & km/s  & Jy km/s &  $10^{7}L_\odot$ & $\rm K\,km/s\,pc^{2}$ & $10^{10}M_\odot$  }
\startdata
MIPS429           & 565$\pm$80           & 777     &  $1.00\pm0.17$   &  3.56    &  2.9    &  2.32   \\   
MIPS506           & 300$\pm$80           & 330     &  $0.52\pm0.11$    &   2.16   & 1.63   &  1.30  \\   
MIPS506$^{dbl}$   & 139$\pm$50,88$\pm$40 & 150,87  &  $0.52\pm0.11$   &   2.16    &  1.63   &  1.30  \\
MIPS8196          &  ...                 & ...     &  $<$0.5          &  $<$2.2  & $<$1.67 &  $<$1.34   \\   
MIPS8327          & 253$\pm50$           & 303     &   $0.42\pm0.07$  &  1.69    &  1.28   &  1.02   \\   
MIPS8342          & 325$\pm70$           & 406     &   $0.65\pm0.07$  &  0.88   &  2.24    &  1.79   \\   
MIPS8342$^{dbl}$  & 134$\pm40$,144$\pm40$ & 128,130  &   $0.65\pm0.07$ & 0.88   &  2.24    &  1.79   \\ 
MIPS15949         & 500$\pm$117           &  129   &  $1.07\pm0.11$    &  3.59  &  2.72    &  2.18   \\      
MIPS16059         & 471$\pm$54            & 426    &  $0.56\pm0.11$    &  2.33    &  1.76  &  1.41   \\   
MIPS16080         & 130$\pm$60            & 159    &  $0.85\pm0.15$    &  2.69    &  2.03  &  1.62   \\   
MIPS16144         & 744$\pm$217           & 824    &  $1.45\pm0.27$     &   4.99   &  3.77  &  3.02   \\   
MIPS16144$^{dbl}$ & 404$\pm$70,404$\pm$70 & 405,334  &  $1.45\pm0.27$   &  4.99    &  3.77  &  3.02   \\

\enddata
\tablenotetext{a}{The sources with double peak CO spectra have two entries per object in this table. The entry marked with $dbl$ has the velocity FWHM fitted with double gaussian profiles, thus the second column FWHM1 has two values per row.}
\tablenotetext{b}{FWHM1 is from the fitting with one or two gaussians,  see above note. }
\tablenotetext{c}{FWHM2 is directly measured from the data as the full width at the half peak intensity, without assuming the gaussian profile.  For the double peak CO spectra,  FWHM2 are measured for both peaks separately.}
\tablenotetext{d}{Here most CO line intensity and luminosity are for J(3-2) transition, except MIPS8342, which is J(2-1).  To derive H$_2$ mass,  we assume $T_{co(3-2)}/T_{co(1-0)}$\,=\,1 and also $T_{co(2-1)}/T_{co(1-0)}$\,=\,1.  $\alpha$ is CO J(1-0) luminosity to H$_2$ (+He) molecular mass conversion factor.  We adopt $\alpha$\,=\,0.8\,$\rm M_\odot$/[K\,km/s\,pc$^2$] for this paper, which is suitable for SMGs and local ULIRGs.}
\end{deluxetable*}

At the far-IR, four sources are detected in the 70\um\ band, and  only one at 160\um.  With so few far-IR detections, is the starburst-component really required?  The only sources for which the answer is obviously yes are MIPS8342 (based on its 160\um\ detection) and MIPS16144 (as it is a strong-PAH/mm-bright source). For MIPS16080, its 70\um\ detection is difficult to explain without such a component, unless we assume a somewhat cooler AGN template.  In particular, the stacked 1.2\,mm detection (1.06\,$\pm$\,0.18\,mJy) requires a minimum amount of star formation, as indicated by the quiescent spiral galaxy template shown as the solid pink line in Figure~\ref{sed} (this is a Sd template from Polletta et al. (2007)). 

This naturally begs the question whether the starburst or the quiescent Sd template is best  suited to fit the SEDs of our sources.  The choice will greatly impact the calculated far-IR luminosity.  We argue that in order to fit simultaneously the observed 70, 160\um\ and 1.2\,mm fluxes, the starburst template is required.  Sd templates can explain the stacked flux at 1.2\,mm, but fail significantly with 70\um\ detections and mid-IR spectra.  
The stacked limit at 70\um\ is (2\,$\pm$\,0.68)\,mJy, high enough that Sd template would fail to fit this limit by a large margin. Definitive solution to this problem will soon to be available from Herschel data with better spatial resolution and sensitivity. 
The luminosities derived from our SED fitting are tabulated in Table~\ref{ancillary2}. 
We note that $\rm L_{FIR}$ is dominated by starburst component for all sources in this study.

\begin{deluxetable*}{lccccccc}
\singlespace
\tablecolumns{7} 
\tablewidth{0in}
\tablecaption{Derived parameters from the ancillary data for the targets \label{ancillary2}}
\tabletypesize{\footnotesize}
\tablehead{
\colhead{ID} &
\colhead{$\rm Log_{10}[L_{IR}^{a}]$} & 
\colhead{$\rm Log_{10}[L_{FIR}^{a}]$} & 
\colhead{R$_{eff}$}  & 
\colhead{EW(7.7PAH)$_{rest}$} & 
\colhead{$\tau_{9.8\mu m}$} & 
\colhead{$\rm Log_{10}[L_{1.4GHz}^b]$} \\
  & L$_\odot$ & L$_\odot$ &  kpc &  $\mu m$    &   &   W\,Hz$^{-1}$     } 
\startdata
MIPS429   & 12.72 & 12.50$\pm0.15$ &  ... & $0.41\pm0.1$  & $>$7.3      & $<24.4$        \\
MIPS506   & 12.93 & 12.70$\pm0.09$ & 1.97 & $0.87\pm0.36$ & $>$6.7      & $24.67\pm0.14$ \\
MIPS8196  & 12.99 & 12.45$\pm0.15$ & 3.31 & $0.31\pm0.08$ & $1.3\pm0.4$ & $<24.5$         \\
MIPS8327  & 12.83 & 12.47$\pm0.21$ & 2.07 & $<0.3$        & $2.4\pm0.7$ & $25.88\pm0.02$  \\
MIPS8342  & 12.55 & 12.13$\pm0.30$ & 2.28 & $0.58\pm0.14$ & $0.2\pm0.4$ & $24.82\pm0.08$  \\
MIPS15949 & 12.90 & 12.42$\pm0.18$ & 2.36 & $0.31\pm0.06$ & $0.0\pm0.1$ & $24.60\pm0.20$  \\
MIPS16059 & 12.87 & 12.57$\pm0.15$ & 2.82 & $0.29\pm0.07$ & $2.7\pm0.8$ & $25.23\pm0.04$   \\
MIPS16080 & 12.71 & 12.38$\pm0.15$ & 2.43 & $0.11\pm0.05$ & $2.1\pm0.5$ & $24.88\pm0.05$   \\
MIPS16144 & 12.72 & 12.61$\pm0.12$ & 3.24 & $2.50\pm0.26$ & $2.0\pm1.4$ & $24.46\pm0.43$    \\

\enddata
\tablenotetext{a}{$\rm L_{IR}$ is integrated over 8\,--\,1000$\mu m$, $\rm L_{FIR}$ is over 40-120$\mu m$. }
\tablenotetext{b}{$\rm L_{1.4GHz}$ is the monochromatic luminosity at the rest-frame 1.4GHz, derived based on the observed radio fluxes at 1.4\,GHz and 650\,MHz.}
\end{deluxetable*}

\section{Results \label{sec:results}}

\subsection{Detections of CO Line Emission}

\begin{figure*}[!t]
\centering{\includegraphics[width=2.5\columnwidth,angle=90]{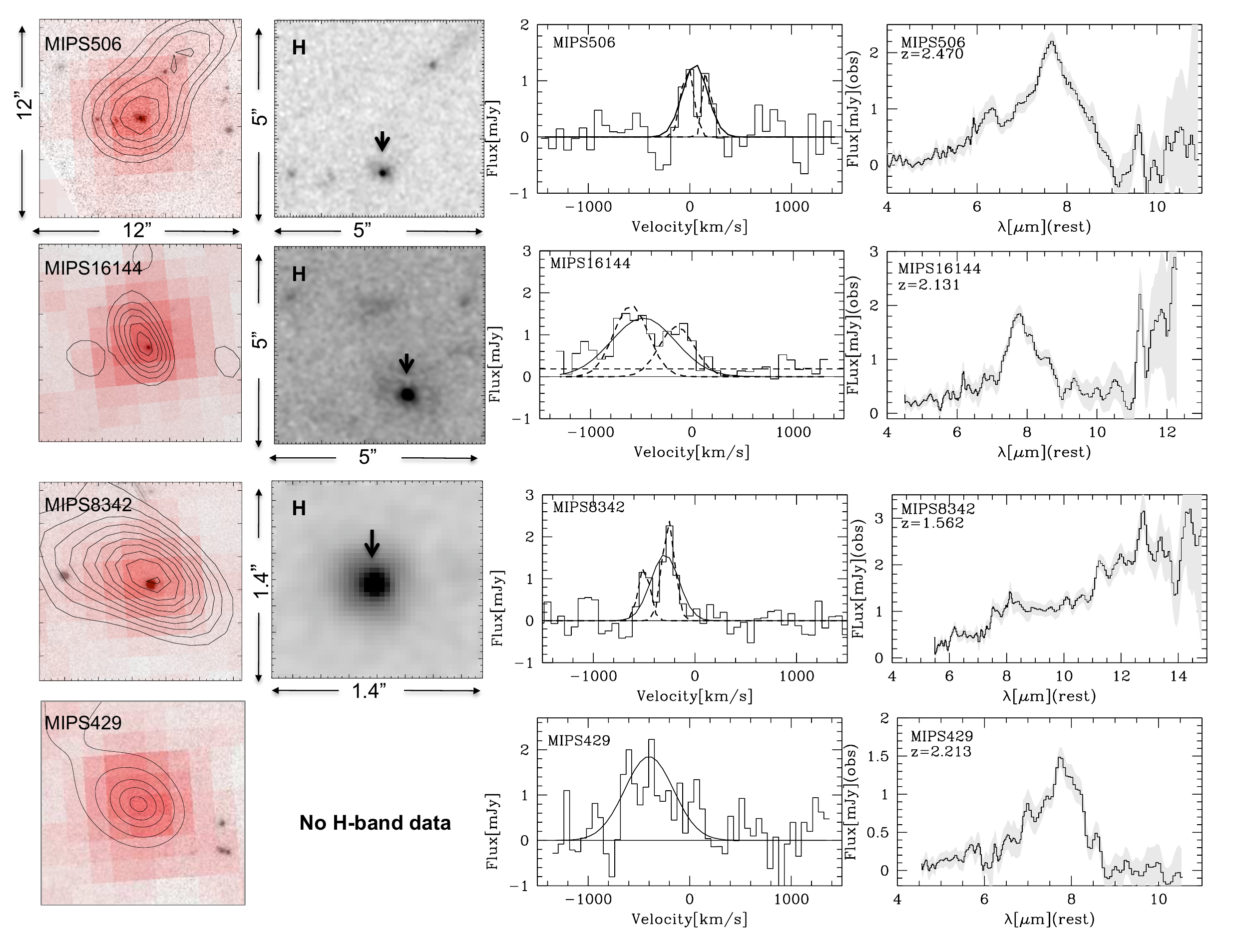}}
%
\caption{\footnotesize The first column shows the the 24\um\ image in pink, HST/NICMOS H-band image in black in its original spatial resolution, and CO image in contours.  In all panels, the CO contours start at 1$\sigma$ with a step size of 1$\sigma$.  For $\sigma$ values, see the text for detail.  The image size is 12$^{''}$$\times$12$^{''}$. The second column shows the H-band morphology.   The third and fourth column show the CO and mid-IR spectra.  The MIPS16144 CO spectrum shows a detection of continuum, marked by the horizontal dashed line. \label{map}}
\end{figure*}

Of the nine sources observed,  eight yielded significant detections in the CO J(3-2) or J(2-1) transitions.  The CO J(3-2) spectrum of MIPS16144 also shows a weak continuum at the level of 0.5\,$\pm$\,0.1\,mJy ($\lambda_{\rm{obs}}$\,=\,2.717\,mm).  This observed continuum flux is consistent with the full-IR SED (see the open triangle symbol in Figure~\ref{sed}).  Figure~\ref{map} presents the CO maps,  HST/NICMOS H-band images and mid-IR spectra, one row per source.  The first column shows the CO contours overlaid on the 24\um\ image (red, FWHM$\sim$\,6$^{''}$) and the {\it HST} NICMOS H-band image (dark, FWHM$\sim$\,0.22$^{''}$).  All CO contours start at $1\sigma$ with an increment contour step size of $1\sigma$. MIPS429 has no NICMOS H-band data, and we used the available WFPC2 F814W image instead.   It has bright 24\um\ flux  but no optical counterpart  in both ground-based $R$ image and the HST/WFPC2 image.  The second column shows the HST/NICMOS H-band images of our sources, providing morphologies at the rest-frame $\sim$\,5500\AA.  The third and fourth columns are the CO and mid-IR spectra.  The mid-IR spectra for MIPS16059 and MIPS16080 have higher SNR, and are from an ultra-deep \spitz observation aiming to detect water ice and hydrocarbons among a sample of deeply embedded $z$\,$\sim$\,2 ULIRGs \citep{sajina09}.

The strength of PAH emission and the 3\,--\,6\um\ hot dust continua are indicators of star formation and AGN respectively.  One interesting question is whether the  mid-IR star formation indicator is correlated with cold dust and molecular gas. Figure~\ref{map} shows one obvious result, that is the detectability of CO gas is not strongly correlated with the mid-IR spectral properties,  with the caveat that this finding is based on a small number of sources.  Of the nine sources,  seven have mid-IR spectra whose prominent features is the broad emission bump at (7-8)\um\ and/or fully or partially observed silicate absorption trough.  This type of spectra is very characteristic for highly obscured 24\um\ sources at $z$\,$\sim$\,2, as shown by several recent \spitz observations \citep{houck05,yan05,yan07}.   Local examples with this type of spectrum are IRAS F00183-7111 and NGC4418 \citep{spoon04a,spoon04b, spoon01}.  One characteristics of these deeply embedded local ULIRGs is their multi-temperature ISM, ranging from hot gas near the central nuclei to cold gas to outer region \citep{spoon04b}.  Our CO observations show that six of these seven sources with deep silicate absorption have cold molecular gas, one (MIPS8196) does not have CO detection at $I_{CO} \simlt 3\sigma$\,$\sim$\,0.5\,Jy\,km/s. 

The CO detection rate of our sample is high, likely due to the improved sensitivity and larger bandwidth of PdBI in recent years. Physically, such high detection rate implies that among 1mJy or brighter 24\um\ ULIRGs at $z$\,$\sim$\,2, a good fraction of them, including SB/AGN composites, deep silicate absorption systems, and AGN dominated galaxies, could have abundant cold molecular gas.   For these 24\um\ bright ULIRGs with substantial black holes in their centers, one important yet unsolved question is how cold molecular gas is exactly turned into stars and feeds the growth of black holes.  The answers will come from future high resolution studies with ALMA.

\addtocounter{figure}{-1}
\begin{figure*}[!t]
\centering{\includegraphics[width=2.5\columnwidth,angle=90]{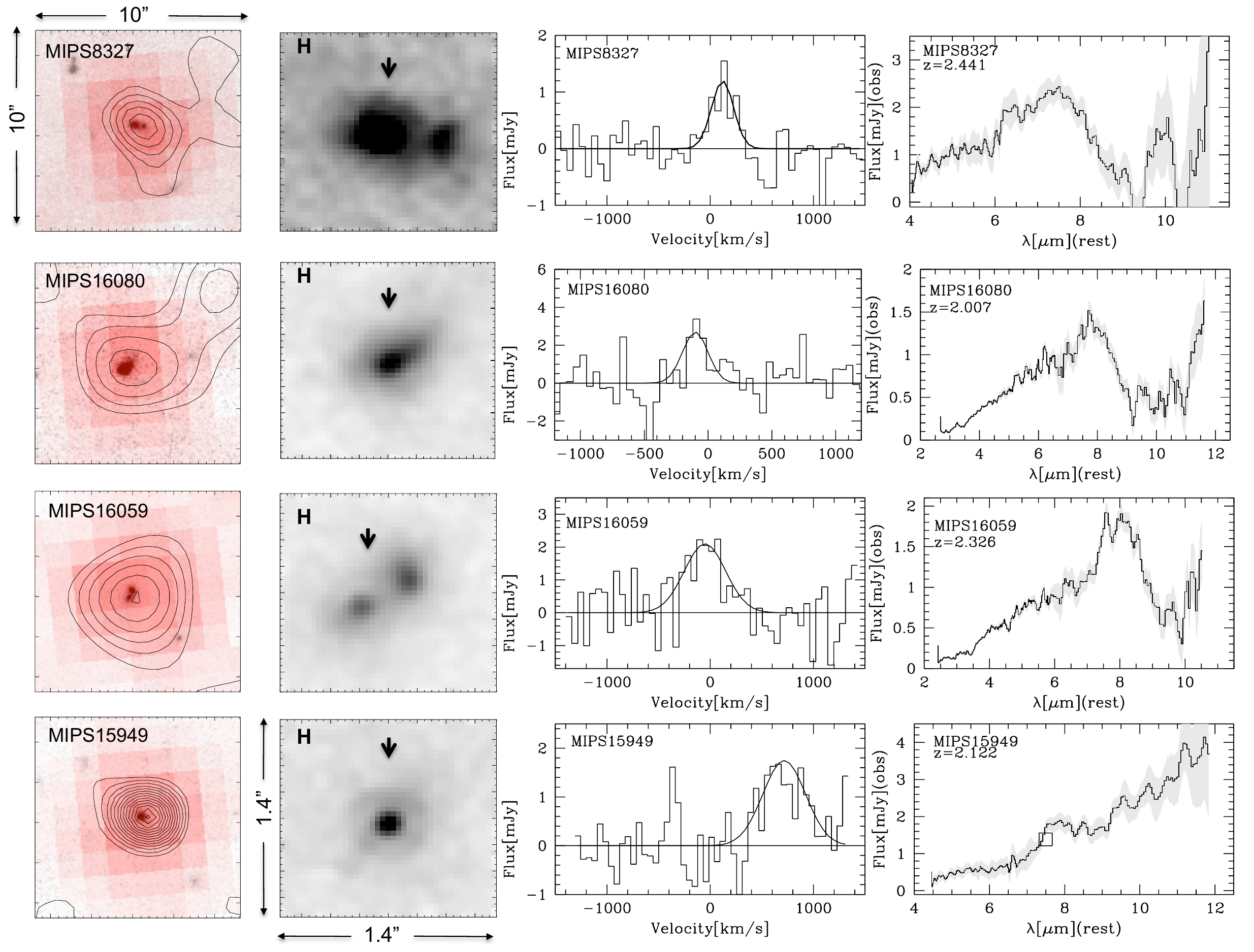}
\caption{\footnotesize Continue. The first column image size is 10$^{''}$$\times$10$^{''}$, and the second column image size is 1.4$^{''}$$\times$1.4$^{''}$. }}
\end{figure*}

\addtocounter{figure}{-1}
\begin{figure*}[!t]
\centering{\includegraphics[width=2.1\columnwidth,angle=0]{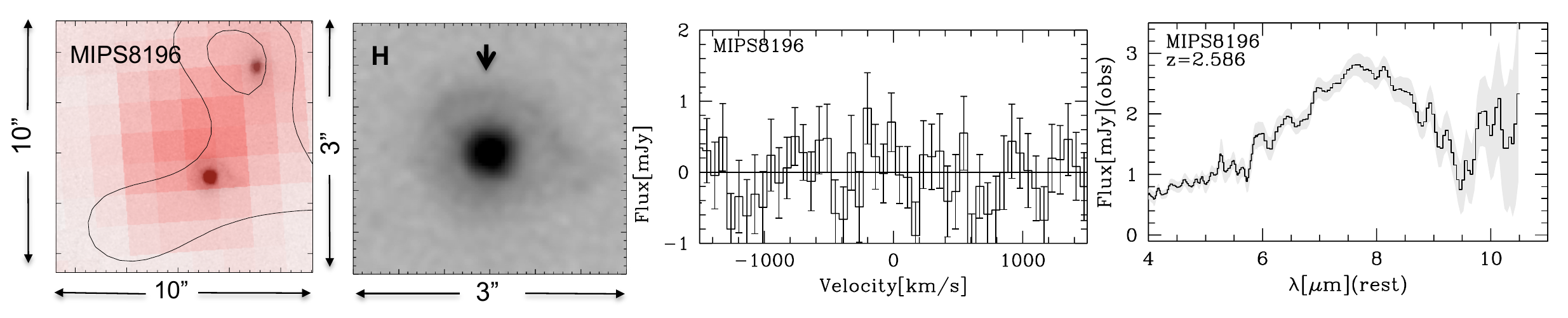}}
\caption{\footnotesize Of 9 objects observed, this is the only source with no significant detection.}
\end{figure*}

\subsection{Cold molecular gas fraction and far-IR emission}

\begin{figure}[!h]
\centering{\includegraphics[width=1.\columnwidth,angle=0]{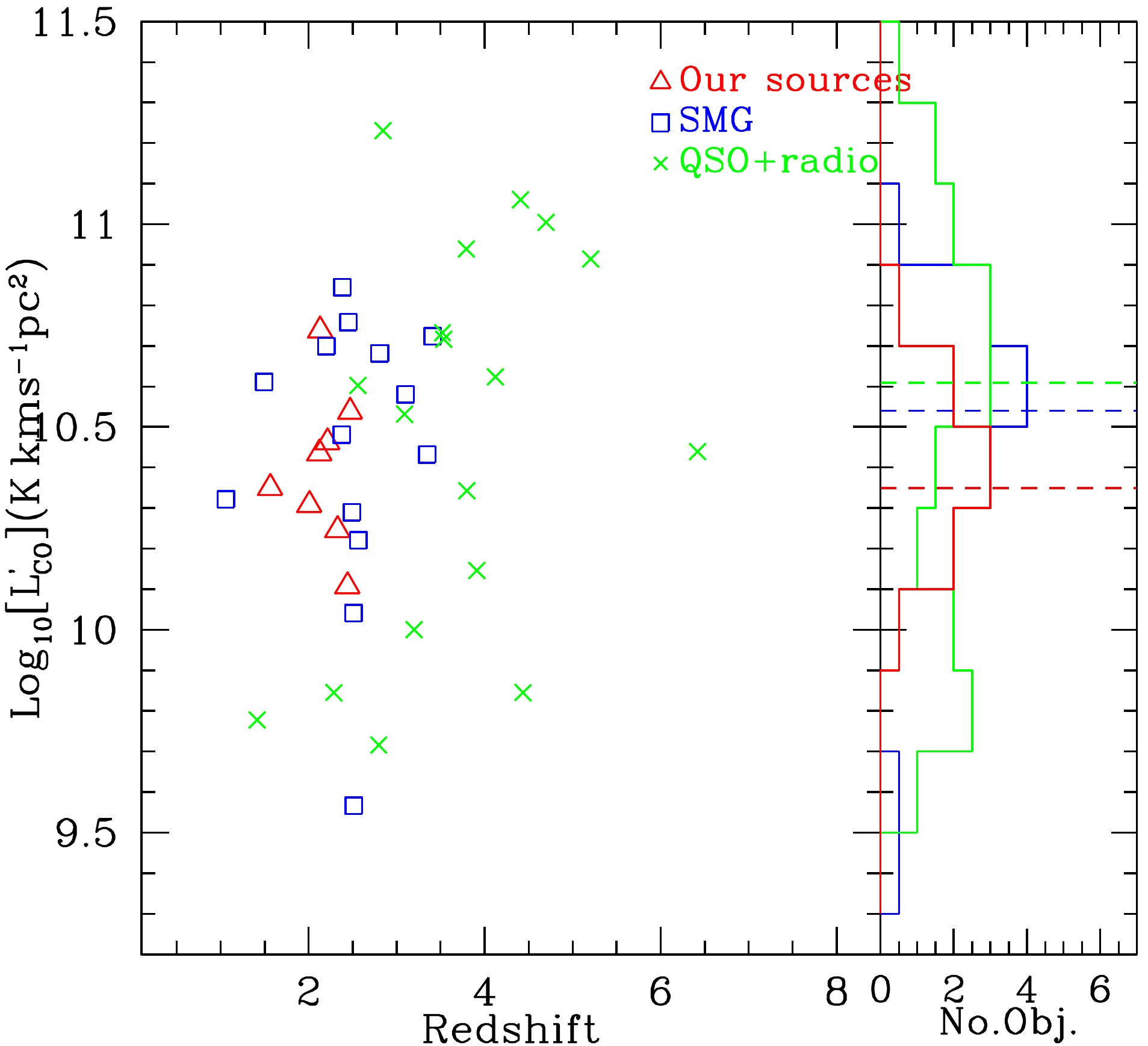}}
%
\caption{\footnotesize This compares the CO integrated line luminosity as a function of redshift for various sources at high-$z$. The SMG data are from Greve et al. (2005) and Tacconi et al. (2006).  The QSO \&\ radio galaxy data are from the compilation of Solomon \&\ vanden Bout (2005).  The dashed lines in the $L^{'}_{CO}$ distribution plot (right panel) mark the median values for the three galaxy populations. The comparison reveals tentative evidence that on average 24\um\ ULIRGs may have less cold molecular gas than those of SMGs and QSOs. \label{lcoz}}
\end{figure}

Table~\ref{coparam} lists the integrated CO line intensity $I_{co}$ in Jy\,km/s and luminosity $L^{'}_{CO}$ for J(3-2) or J(2-1) transition.  We emphasize again that to derive cold molecular gas mass, we assume $T_{co(3-2)}/T_{co(1-0)}$\,=\,1,  $T_{co(3-2)}/T_{co(1-0)}$\,=\,1 and  CO J(1-0) luminosity to molecular gas mass conversion factor $\alpha$\,=\,0.8\,$M_\odot$/[K\,km/s\,pc$^2$].  The eight sources have $\rm M_{gas}$\,$\sim$\,(1.02\,--\,3.02)$\times10^{10}M_\odot$, with a median value of 1.7$\times10^{10}M_\odot$.  In order to compare our sample with other galaxy populations at $z\simgt 2$ with CO detection, we compiled a list of SMGs based on the published data \citep{greve05, tacconi06,tacconi08}, shown in Table~\ref{smg}.  This sample of 14 SMGs includes only sources with significant detections, and their median gas mass $\rm M_{gas}$ is 3.05$\times10^{10}M_\odot$.  In addition, a sample of 19 QSOs and radio galaxies with low resolution CO observations \citep{Solomon05} is also used as a QSO comparison sample.  As described in \S\,\ref{mass}, we have also constructed the second QSO comparison sample, much smaller subset but with reliable dynamical masses from the recent high resolution CO observations. Figure~\ref{lcoz} left panel shows $L^{'}_{CO}$ versus redshift for the 24\um\ ULIRGs and the comparison samples of SMGs, QSOs and radio galaxies.  The right panel in Figure~\ref{lcoz} presents the histogram of $L^{'}_{CO}$ with the dashed lines marking the median values of the three populations. 
The average CO mass of our 24\um\ ULIRG sample is about a factor of 2 less than the average values calculated from the specific comparison samples of SMGs and gas rich QSOs currently available in the literature. Kolmogorov-Smirnov tests on 24\um\ ULIRGs versus SMGs and 24\um\ ULIRGs versus QSOs derived the probabilities of two datasets drawn from the same parent sample of 0.19 and 0.189 respectively.  We emphasize that the statistics is still too small to draw any definitive conclusions for the overall populations of these three types of sources. 


\begin{figure}[!h]
\centering{\includegraphics[width=1.\columnwidth,angle=0]{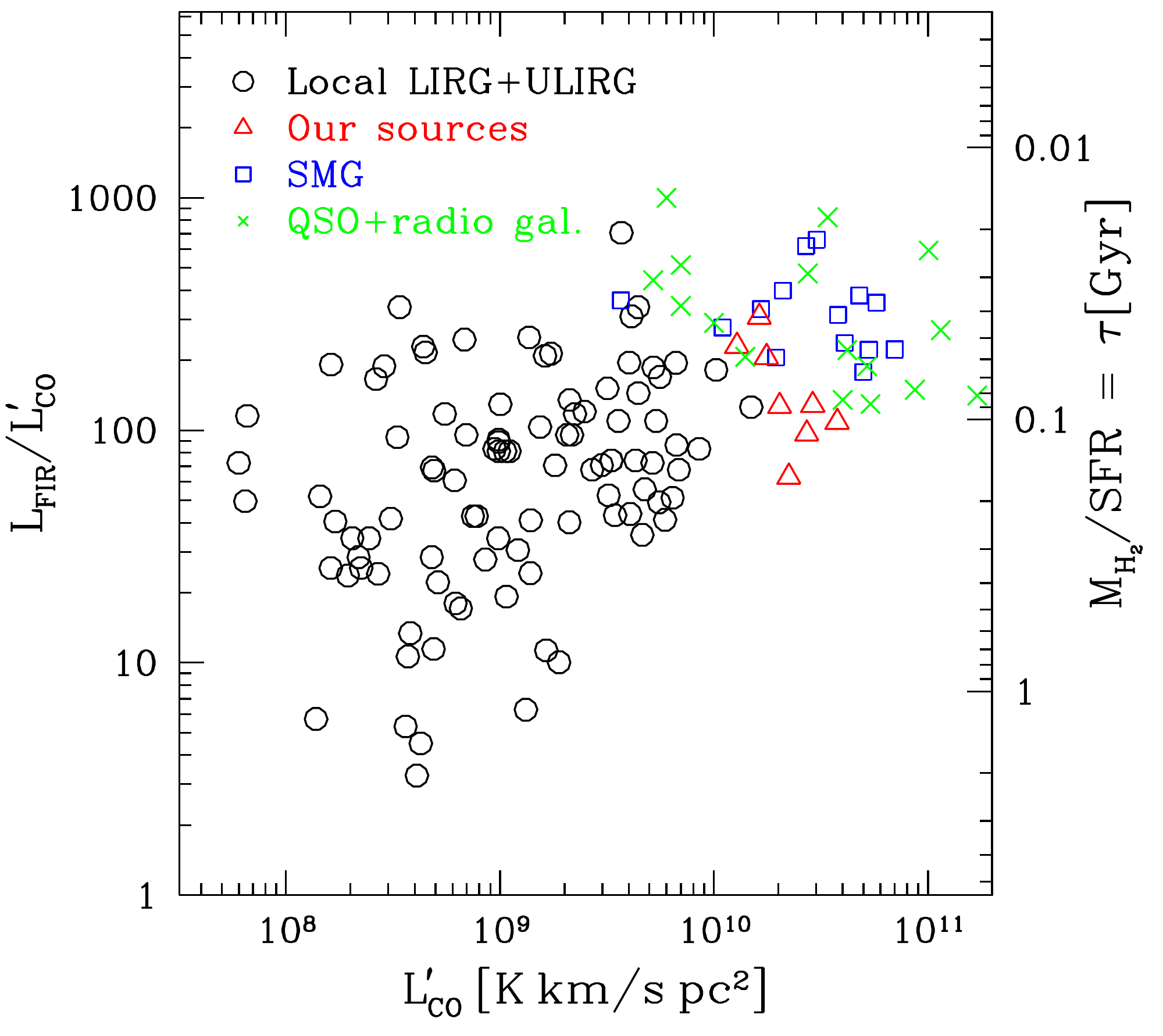}}
%
\caption{\footnotesize The plot shows the ratio of $\rm L_{FIR}[L_\odot]$ and $\rm L^{'}_{CO}[K\,km/s\,pc^2]$ versus $\rm L^{'}_{CO}$.  The black, open circles indicate local infrared and ultra-luminous galaxies from \citet{solomon97}, the red, open triangles are our data points,  the blue, open squares  the SMGs the published data from \citet{greve05,tacconi06,tacconi08},  and the green crosses the QSO \&\ radio galaxy data compiled by \citet{Solomon05}. \label{lfirlco}}
\end{figure}

If far-IR emission is considered mostly from massive young stars born in recent starbursts, $\rm L_{FIR}$ can be directly related to star formation rate (SFR).  Therefore,  the ratio of $\rm L_{FIR}/L^{'}_{CO}$ can be interpreted as describe (1) gas depletion time scale, $\tau_{SF}$\,=\,${\rm M_{gas}/SFR}$ or (2) star formation efficiency, $\rm SFE$\,=\,$\rm L_{FIR}/M_{gas}$, {\it i.e.} how much luminosity can be produced by 1\,$M_\odot$ of gas.  If we take the $\rm SFR[M_\odot/yr]$\,=\,$\rm c*\bigl({L_{FIR} \over 10^{10}L_\odot}\bigr)$, where c\,=\,0.8\,--\,3, depending on the definition of $\rm L_{FIR}$ \citep{kennicutt98, meurer97}.  We adopt c\,=\,1.5 because we use $\rm L_{FIR}$\,=\,$\rm L_{40-120\mu m}$ and the commonly adopted relation by Kennicutt (1998) has c\,=\,1.72 with $\rm L_{FIR}$\,=\,$\rm L_{8-1000\mu m}$.  The $\rm L_{FIR}$ is measured from the full SED fitting described in \S~\ref{sec:sed}.  The inferred $\rm SFR$ are 32\,--\,787\,M$_\odot$ yr$^{-1}$, with the mean value of $585\,M_\odot$ yr$^{-1}$.  These values are substantially larger than normal star forming galaxies, but a factor of 2\,--\,3 less than what has been observed for SMGs at the similar redshifts.  Furthermore, if we assume that 70\%\ of $\rm L_{FIR}$ from SMGs is due to star formation, as suggested by some recent studies \citep{dave08},  the averaged SFR for SGMs would still be higher than that of 24\um\ sources by a factor of (1.5\,--\,2).   The inferred values for $\rm SFE$ and gas depletion time scale $\tau_{SF}$ range from 69\,--\,388\,$L_\odot\,M_\odot^{-1}$($\langle SFE \rangle$\,=\,330\,$L_\odot \,M_\odot^{-1}$) and 17\,--\,96\,Myr ($\langle \tau_{SF}\rangle$\,=\,38\,Myrs) respectively.   For comparison,  star formation efficiency $\langle SFE \rangle$ is typically $(180\pm160)L_\odot M_\odot^{-1}$ for local ULRIGs \citep{solomon97} and $450\pm170\,L_\odot M_\odot^{-1}$ for SMGs \citep{greve05,tacconi06}.   And the gas depletion time scale $\tau_{SF}$ ranges from 10\,--\,100\,Myrs for SMG, and a factor of 10 longer for local LIRGs and ULIRGs \citep{Solomon05}.  Figure~\ref{lfirlco} displays the $\rm L_{FIR}/L^{'}_{CO}$ ratio versus $\rm L^{'}_{CO}$ (or equivalently cold molecular gas), visually illustrating our conclusion that these bright 24\um\ selected $z$\,$\sim$\,2 ULIRGs have less cold molecular gas and smaller star formation efficiency than that of SMGs.   

\subsection{CO velocity widths and dynamical mass estimates \label{mass}}

\begin{deluxetable*}{lcccccc}
\singlespace
\tablecolumns{7} 
\tablewidth{0in}
\tablecaption{Comparison Sample I -- SMG data \label{smg}}
\tabletypesize{\footnotesize}
\tablehead{
\colhead{ID} &
\colhead{$z_{co}$} &
\colhead{L$_{FIR}$} & 
\colhead{L$^{'}_{CO(3-2)}$$^a$} & 
\colhead{V$^b_{FWHM}$} & 
\colhead{double}  &
\colhead{Ref.$^c$}  \\
  &  & L$_\odot$ & K km/s pc$^2$ & km/s &     }
\startdata
SMMJ023956$-$0134 &  1.0620  &  8.36e+12 &  2.1e+10   &  $780\pm60$  & yes & (1) \\
SMMJ023951$-$0136 &  2.8076  &  1.82e+13 &  4.8e+10   & $1360\pm50$ & yes & (1)  \\
SMMJ044307+0210 &  2.5094  &  3.04e+12 &  1.1e+10   &  $350\pm60$ &  yes  & (1)   \\
SMMJ094303+4700 &  3.3460  &  1.67e+13 &  2.7e+10   & $420\pm50$ &  yes  & (1)    \\
SMMJ123549+6215(HDF76) &  2.2021  &  8.90e+12 &  5.0e+10   &  $600\pm50$ & yes & (2)  \\ 
SMMJ123707+6214(HDF242) &  2.490  &  4.00e+12 &  2.0e+10   & $430\pm60$ & yes &  (2)  \\
SMMJ131201+4242 &  3.408  &  1.18e+13 &  5.3e+10   &  $530\pm50$ &  no & (1)  \\
SMMJ140103+0252 &  2.5653  &  5.51e+12 &  1.7e+10   &  $200\pm40$ & no  & (1)   \\
SMMJ163554+6612 &  2.5168  &  1.33e+12 &  3.7e+9   &  $500\pm100$ & yes & (1)  \\
SMMJ163650+4057(N2850.4) &  2.3853  &  1.56e+13 &  7.0e+10   &  $840\pm110$  & yes & (2) \\
SMMJ163658+4105(N2850.2) &  2.4500  &  2.03e+13 &  5.7e+10   & $870\pm80$ & yes & (2)  \\
SMMJ163706+4053 &  2.3800  &  2.00e+13 &  3.0e+10   &  $830\pm130$ & no & (1)  \\
EROJ164502+4626 &  1.4950  &  9.69e+12 &  4.1e+10   &  $400\pm20$  & no & (1)  \\
SMMJ221735+0015 &  3.0990  &  1.20e+13 &  3.8e+10   &  $780\pm100$ & no & (1)  \\

\enddata
\tablenotetext{a}{All \citet{greve05,tacconi06,tacconi08} assume that $T_{co(3-2)}/T_{co(1-0)}$\,=\,1 and $\alpha$\,=\,0.8$M_\odot$/[K,km/s\,pc$^2$] for the molecular gas mass calculation.}
\tablenotetext{b}{The line velocity width is from a single gaussian fit to the spectral data.}
\tablenotetext{c}{References: (1) Greve et al. 2005, MNRAS, 359, 1165.   This paper gives the original references for some of the sources which were not observed by  Greve et al. 2005.  (2) Tacconi et al. 2008, ApJ, 680, 246}
\end{deluxetable*}

\begin{figure}[!h]
\centering{\includegraphics[width=1.\columnwidth,angle=0]{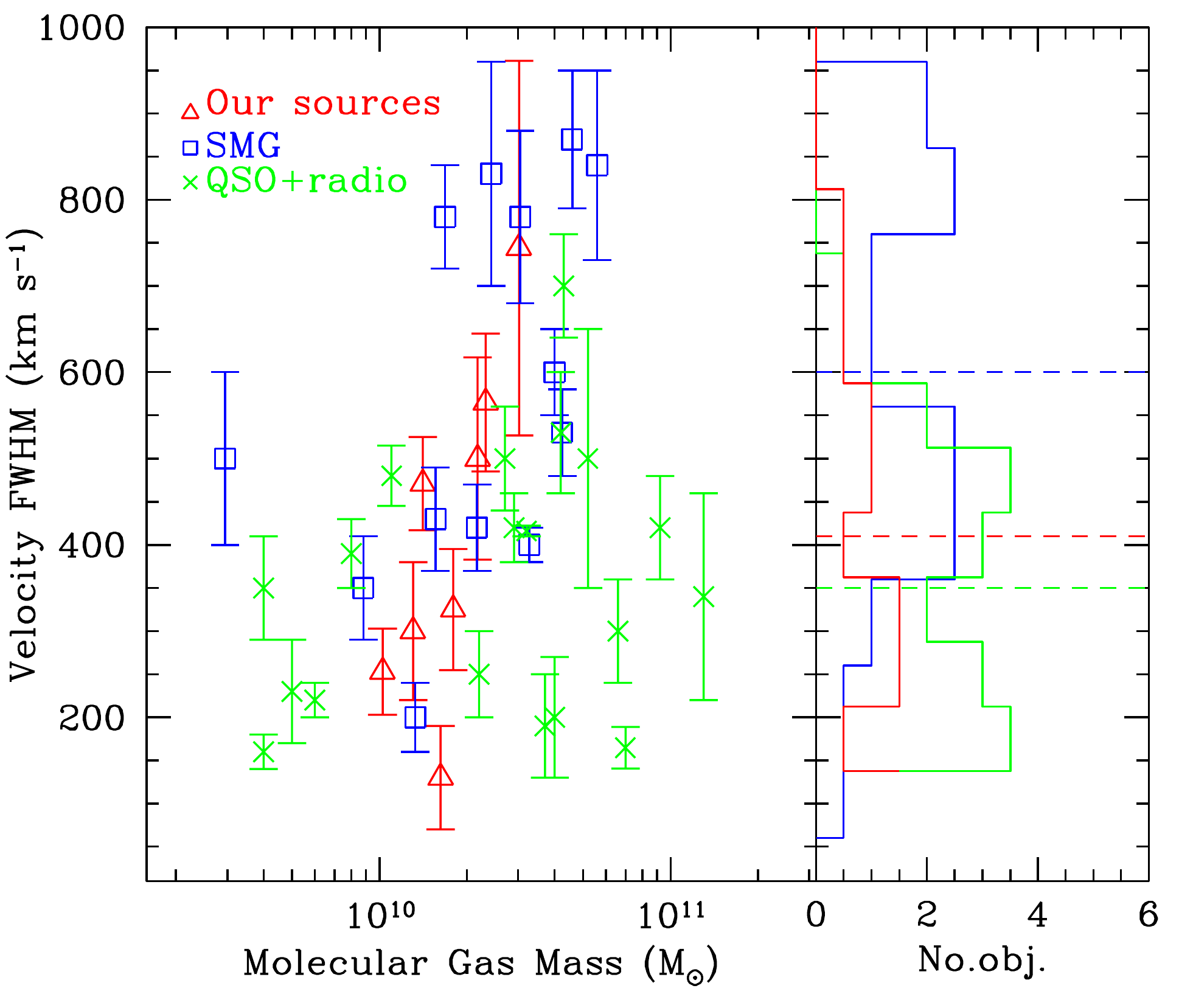}}
%
\caption{The plot compares the CO velocity width (FWHM) versus molecular gas masses of our sources with that of high-z QSO \&\ radio galaxies and SMGs.  All velocity widths are based on a single component fitting.  The SMG comparison sample is listed in Table~\ref{smg}, compiled from \citet{greve05,tacconi06}, and the QSO \&\ radio galaxy data are from the compilation of  \citet{Solomon05,greve05}.  The right panel shows the velocity distributions of three populations with the dashed lines marking the median values. The velocity width distributions show tentative evidence that 24\um\ ULIRGs are similar to QSOs, having smaller CO velocity width than the averaged value for SMGs. \label{velmass}}
\end{figure}

The CO velocity width (FWHM) can be measured by using either a single or double gaussian fit or directly from the data (full width at the half peak intensity).  Table~\ref{coparam} shows the results using both methods.  Assuming a single component and without profile fitting (Table~\ref{coparam} FWHM2),  we have velocity widths ranging from 128\,km/s to 824\,km/s, with a median value of 406\,km/s.  If we use two velocity components for sources with double peak profiles,  the averaged velocity width reduces to 275\,km/s.  With a single component fitting,  Figure~\ref{velmass} compares the CO velocity widths of our sources with that of SMGs and QSOs.  Here the SMG comparison sample is listed in Table~\ref{smg} for 14 objects compiled from \citet{greve05, tacconi06,tacconi08}.  The QSO comparison sample is compiled by \citet{Solomon05} (also see \citet{greve05}).  The right panel presents the velocity distributions, with dashed lines marking the median values of 406, 635 and 350\,km/s for \spitz $z$\,$\sim$\,2 ULIRGs, SMGs and QSOs respectively.  As the local reference,  the $z$\,$\sim$\,0 ULIRGs have a median CO velocity width of 240\,km/s \citep{sanders91,solomon97}. 
Kolmogorov-Smirnov tests on these three small samples produce the probabilities of 0.23 and 0.84 respectively, for $\rm \Delta V(FWHM)$ of 24\um\ ULIRGs versus SMGs and 24\um\ ULIRGs versus QSOs drawn from the same parent distributions.



Recent N-body/Smoothed-Particle Hydrodynamic (SPH) numerical simulation combined with 3D polychromatic radiative transfer by \citet{desik09} show that SMGs are mostly produced by major mergers of gas rich, star forming dominated progenitors,  whereas 24\um\ bright ULIRGs at $z$\,$\sim$\,2 could be formed by both major and minor mergers. This model predicts that CO velocity widths of bright 24\um\ ULIRGs spread over (100-1500)\,km/s, overlapping with that observed among SMGs, but the median value is smaller than that of SMGs.  



Spatial distribution of molecular gas holds the imprints of kinematic state of a galaxy.  Although the majority of our sample is unresolved,  two targets, MIPS506 and MIPS16144, show spatially extended CO emission across regions of $\sim$\,100 and 46\,kpc,  significant after taking into account of their beam sizes of 5.02$^{''}$$\times$4.37$^{''}$$\sim$\,41\,kpc$\times$36\,kpc and 2.44$^{''}$$\times$2.15$^{''}$\,$\sim$\,20\,kpc$\times$18\,kpc respectively.  MIPS506 shows two distinct components with a separation of 45\,kpc.   
The CO spectra in Figure~\ref{map} show that MIPS506, MIPS16144, and MIPS8342 have double peaked velocity profiles, with single and double gaussian fit (solid and dashed lines respectively) to the observed data (histogram).   A double peak CO velocity spectrum is indicative of either a rotating disks or of two galaxies in an ongoing merger. Our CO data measures 3 out of 9 (33$\pm19$\%) sources having double velocity components.  The low resolution CO data for the SMG comparison sample in Table~\ref{smg} shows 64$\pm21$\%\ (9/14) having double or multiple velocity components.  If we consider pure Poisson statistical errors, these two fractions are consistent within $1\sigma$.  



\begin{figure}[!h]
\centering{\includegraphics[width=1.\columnwidth,angle=0]{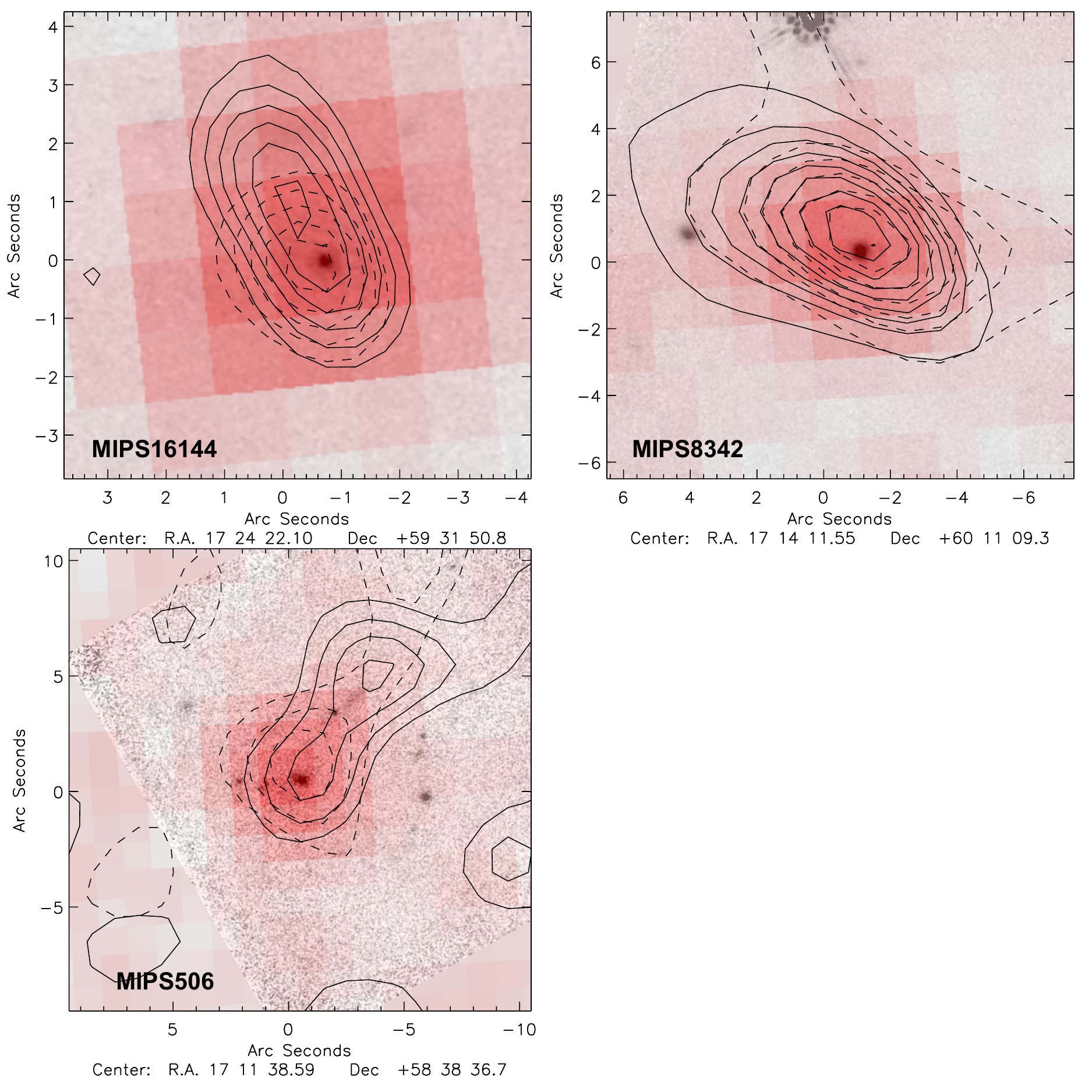}}
%
\caption{\footnotesize Overlaid on the 24\um\ (pink) and NICMOS H-band (black) images are the CO contours of the integrated maps over the positive (solid lines) and negative (dashed lines) velocity peaks for the 3 sources with double peak profiles. The solid and dashed contours are for the CO velocity peaks at (-138, -605)\,km/s, (160,-15)\,km/s and (-250,-520)\,km/s for MIPS16144,MIPS506 and MIPS8342 respectively. The contour starts at 1$\sigma$ with a step of 1$\sigma$ for all three panels. For the detail, see the text in \S\ref{mass}. \label{redblue}}
\end{figure}

One important question is whether the two velocity components in each of the three double peak sources are spatially separated. Figure~\ref{redblue} shows the CO contours overlaid on the 24\um\ (pink) and {\it HST} H-band (black) images, with solid and dashed contours indicating two separate velocity peaks. Here, the integrated CO maps are summed over (4\,--\,5) velocity channels (see Table~\ref{obs} for the channel width), covering the respective velocity peak. For example, the integrated CO maps for MIPS16144 shown in Figure~\ref{redblue} are summed from -730\,km/s to -330\,km/s for the velocity peak at -605\,km/s and from -330\,km/s to -70\,/km/s for the velocity peak at -138\,km/s.  The similar maps are made for MIPS506 and MIPS8342, and shown in Figure~\ref{redblue}.  The rms $\sigma$ values used for the contours are in Table~\ref{obs}. The two velocity peaks in MIPS16144 have a clear spatial separation of $1.3^{''}$, which corresponds to 10.9\,kpc.  The two velocity components in MIPS8342  are spatially too close to be resolved.  In the case of MIPS506, the separation is $\sim5.5^{''}$, corresponding to 45\,kpc.  Such a large distance, particularly for cold molecular gas,  suggests that MIPS506 is probably a pair of merging galaxies,  with substantial dust extinction in the rest-frame optical band.  This explains the very faint multiple blobs (2.8$^{''}$$\sim$\,24\,kpc separation) in the H-band image ($6^{''}\times5^{''}$) shown in Figure~\ref{map}.   We have examined our shallow 3.5\,--\,8\um\ IRAC images to see if the galaxy pair detected in the CO map is also detected in other bands,  only an unresolved object is detected at 3.5 and 4.8\um, and no detections at other two IRAC bands.  The physical model for MIPS16144 and MIPS8342 is probably also merger, although rotating disk model is also consistent.  Additional supporting evidence for \spitz ULIRGs being mergers comes from the HST/NICMOS H-band images, with the detailed discussion in \S\ref{hst}.  



For MIPS16144 and MIPS506 with spatially resolved CO maps, assuming a two-body interaction, we can approximate the total dynamical mass with $\rm M_{dyn}$\,=\,$\rm R$*$\rm v_{circ}^2/G$\,=\,$\rm R$*$\rm \Delta V^2/(sin^2(i) G)$, where $\rm sin(i)$ is the inclination angle, $\rm R$ is the separation between two merging objects, and $\rm \Delta V$ is the observed CO peak velocity difference. $\rm R$ and $\rm \Delta V$ are (10.9\,kpc, 463\,km/s) and (45\,kpc,174\,km/s) for MIPS16144 and MIPS506 respectively.  We derived ${\rm M_{dyn}}$\,=\,5.4$\rm \times 10^{11} sin^{-2}(i)$ and 3.2$\rm \times10^{11} sin^{-2}(i)M_\odot$ for MIPS16144 and MIPS506. If assuming an averaged value of $\langle$sin$^2$(i)$\rangle$\,=\,$2/3$,  the gas fraction, $\rm M_{gas}/M_{dyn}$,  is 4\%\ and 3\%\ for MIPS16144 and MIPS506 respectively.  Although several recent papers have published gas fractions for various types of galaxies, we caution that the equation used for calculating the dynamical masses based on observed emission line FWHM could differ by a factor of (2\,--\,5), see \citet{neri03, erb06, tacconi06, riechers09}.  



\begin{deluxetable}{lccccc}
\singlespace
\tablecolumns{6} 
\tablewidth{0in}
\tablecaption{Comparison Sample II -- QSO data$^a$ \label{qso}}
\tabletypesize{\footnotesize}
\tablehead{
\colhead{ID} &
\colhead{$z_{co}$} &
\colhead{$\rm M_{BH}$} & 
\colhead{$\rm M_{gas}$} & 
\colhead{$\rm M_{dyn}$} & 
\colhead{incl. angle}  \\
  &   & $10^9$      & $10^{10}$   &  $10^{10}$  &    \\ 
  &  & $M_\odot$ & $M_\odot$ & $M_\odot$ &     }
\startdata
APM08279+5255 &	3.91 &	23&	13&	22&	25deg  \\
PSSJ2322+1944 &	4.12&	1.5&	1.7&	4.4&	no corr. \\
BRI1335-0417  &	4.41&	6&	9.2&	10&	no corr. \\
SDSSJ1148+5251&	6.42&	3&	2.2&	5.5&	65deg  \\

\enddata
\tablenotetext{a}{The black hole masses are derived from UV spectroscopy,  the gas and dynamic masses from the high spatial resolution CO observations.  References:  \citet{walter04,riechers08a,riechers08b,riechers09}.}
\end{deluxetable}

 The stellar half-light radii have been measured from the HST/NICMOS H-band (1.6\um) images for our sample \citep{dasyra08,michel09}.  Using $\rm M_{dyn}$\,=\,$\rm R_{1/2}$*$\rm v_{circ}^2/G$\,=\,$\rm R_{1/2}$*$(\rm \Delta V^{CO}_{FWHM})^2/(sin^2(i) G)$ and assuming that cold molecular gas disks are smaller than stellar half-light radii,  we can also set limits on the dynamical masses of additional 5 sources in our sample.  This assumption is usually correct for nearby galaxies \citep{sanders91,solomon97}.  We note that our sources probably have fairly high dust extinction at the observed H-band (rest-frame 5000\AA), which will cause systematic under-estimates of the intrinsic stellar sizes.  Using the half-light radii are listed in Table~\ref{ancillary2},  we obtained $\rm M_{dyn}$\,$\sim$\,(0.2\,--\,2.2)$\rm \times$10$^{11}M_\odot$ for MIPS8327, MIPS8432, MIPS15949, MIPS16059 and MIPS16080,  using an averaged inclination angle.  These rough estimates are on the same order of magnitude in comparison with the averaged dynamical mass of SMGs, 1.2$\times$10$^{11}M_\odot$ \citep{greve05,tacconi06}. 

\subsection{Our targets are likely mergers \label{hst}}
\begin{figure*}[htbp]
\centering{\includegraphics[width=2.\columnwidth,angle=0]{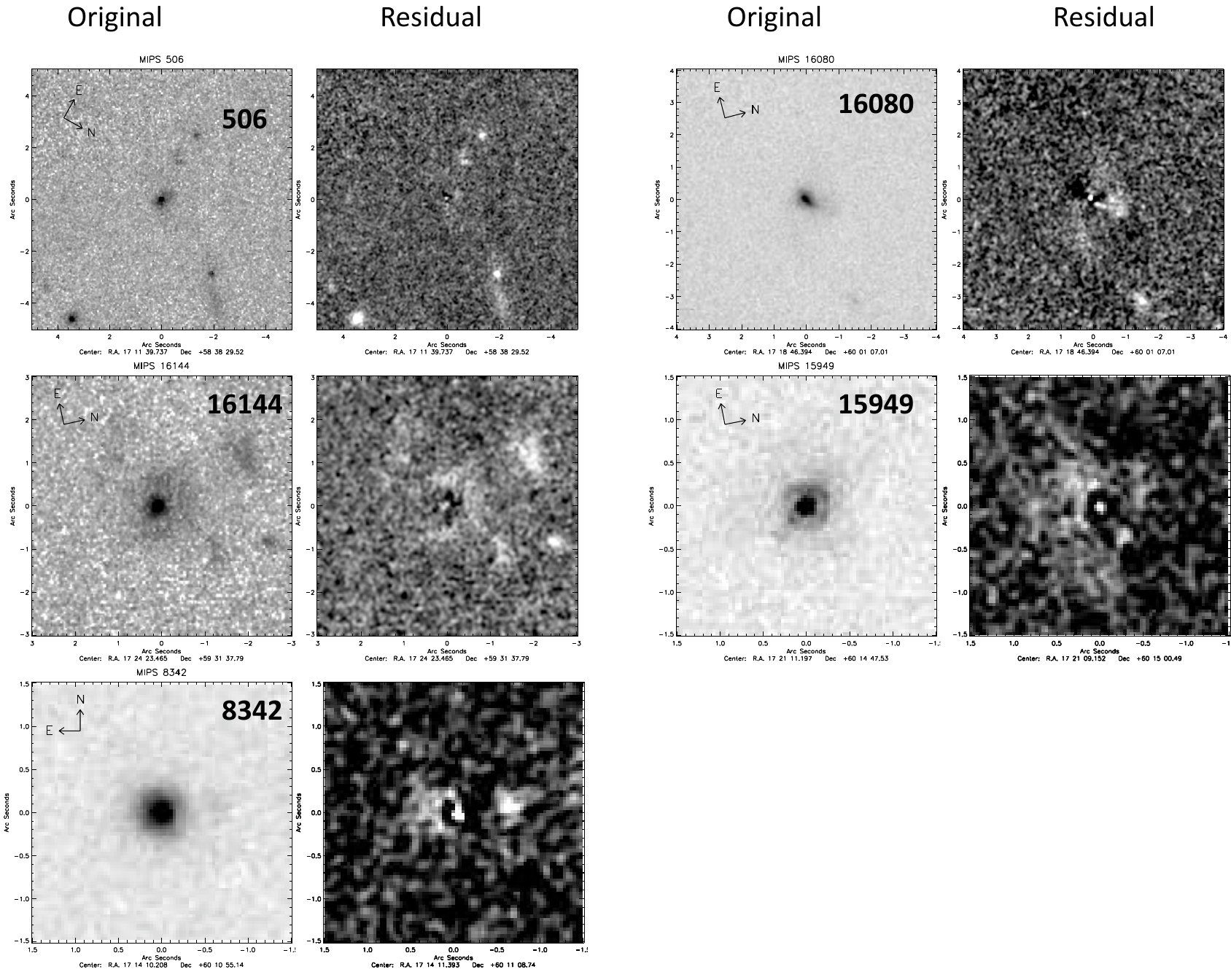}}
%
\caption{\footnotesize The HST H-band images and their residual images with a sersic profile model subtracted for the 5 sources which do not show distinct galaxy pairs.  The intensity scale is reversed for the residual images in order to bring out the faint features. These images reveal morphological signatures suggesting dynamical interactions. \label{merger}}
\end{figure*}

The CO images and spectra in Figure~\ref{comap} and Figure~\ref{map} have revealed that at least one source, MIPS506, is a merger involving two galaxies with substantial cold molecular gas.  In addition,  MIPS8327 and MIPS16059 are likely interacting galaxies because their HST/NICMOS H-band images show close companions within separations of 3.3 and 4\,kpc ($0.4^{''} \&\ 0.5^{''}$) respectively.   \citet{michel09} performed detailed 2-dimensional light profile fitting to the HST images, assuming a sersic profile model\footnote{$I(r)$\,=\,$\rm I_e$*$\rm \exp(-\kappa*[(r/r_e)^{1/n}-1])$, $\kappa$ is a function of sersic index $n$. Sersic index $n$ is 4 for de Vaucouleurs profile and 1 for the exponential profile.}. Figure~\ref{merger} compares the original H-band images with the model subtracted, residual images (reversed intensity scale) for 5 of our sources with CO detections, MIPS506, MIPS16144, MIPS8342, MIPS16080 and MIPS15949.  
Multiple blobs an clear tidal tails in Figure~\ref{merger} suggests that MIPS506, MIPS16144 and MIPS16080 are probably the products of galaxy interactions.  Although the original H-band images of MIPS8342 and MIPS15949 show isolated early type galaxies, but the faint disks/rings in the residual images indicate that they could be late stage mergers.  Majority sources in our CO sample have merger signatures either in their CO maps or in the H-band images.  
Considering the published studies of SMGs based on HST images and integral spectroscopy \citep{chapman03,conselice03,swinbank06}, we speculate that SMGs, bright 24\um\ ULIRGs, and gas rich QSOs are connected by an evolutionary model involving mergers, with SMGs being in an early stage of mergers of two gas rich, star forming progenitors,  and bright 24\um\ ULIRGs with substantial obscured AGN in a later merger stage, while QSOs are fully merged with massive black holes in place, producing  powerful feedback clearing away dust obscuration.  This simple picture does not explain how SMGs and gas rich QSOs could have such a different history of star formation and black hole growth. Specifically, most SMGs have very weak AGNs, and their estimated black hole to stellar mass ratio is roughly on the order of 1$:$10$^4$ \citep{borys05,dave08}, whereas, recent studies (see Table~\ref{qso}) found the same ratio for these $z\simgt3$ QSOs is on the order of 1$:$10 \citep{riechers08a,riechers09}. For our objects, black hole mass is calculated from the AGN bolometric luminosity estimated in the SED decomposition (see Figure~\ref{sed}).  Stellar mass is computed from the rest-frame $\rm L_{H}$ assuming a $\rm M_{stars}/L_{H}$\,$\sim$\,0.48 from a 2\,Gyr SSP template of Maraston (2005).  The inferred $\rm M_{BH}/M_{stars}$ is on the order of 1$:$10$^3$.  Again, this suggests that \spitz ULIRGs may be a transitional type of sources between SMGs and QSOs.  To really understand the physical relation among these three types of sources,  we will need better UV optical spectra, complete SEDs, and high resolution CO data for \spitz ULIRGs.

\section{Summary and discussions \label{sec:sum}}

We report CO interferometric observations of nine $z$\,$\sim$\,2 ULIRGs with  $S_{24\mu m}$\,$\simgt$\,1\,mJy.  
Combining high spatial resolution HST images, \spitz mid-IR spectra, photometry and the CO data,  we find the following results:


\begin{itemize}

\item{Of the nine sources observed at PdBI,  eight sources have significant detections of the CO J(3-2) or J(2-1) transition with intensity $I_{c} \sim$\,(0.5\,--\,1.5)\,Jy\,km\,s$^{-1}$ at (4.7\,--\,9.7)$\sigma$.  CO line emission is detected from sources with a variety of mid-IR spectral type, including strong PAH, deep silicate absorptions and mid-IR power-law.  The observed CO J(3-2) [or (2-1)] line luminosities $L^{'}_{CO}$ are (1.3\,--\,5.5)$\times$10$^{10}$\,K\,km/s\,pc$^2$, yielding cold molecular gas masses of (1\,--\,4.4)$\times$10$^{10}M_\odot$, based on the assumption of $T_{CO(3-2)}/T_{CO(1-0)}$\,=\,1 and the CO (1-0) luminosity to H$_2$ gas mass conversion factor of 0.8\,$M_\odot$/[K\,Km/s\,pc$^2$].   Based on small statistical samples, we find tentative evidence suggesting that \spitz ULIRGs have a factor of 2 less cold molecular gas than what observed among SMGs and gas rich QSOs. }
 
\item{The CO velocity width distribution of \spitz ULIRGs is similar to that of QSOs with CO detections,  but on average a factor of 1.5 smaller than that of SMGs, with caveat that the available CO statistics is small.  
Two objects, MIPS506 and MIPS16144, have spatially resolved CO emission across regions of 100 and 46\,kpc respectively, significantly extended after taking into account of their beam sizes and shapes.  
Three of our nine targets (33\%) have double peaked CO spectra, in comparison with 64\%\ observed among SMGs.  Two of these three, MIPS506 and MIPS16144,  have their two velocity peaks spatially separated by $\sim$\,45\,kpc and 10.9\,kpc respectively.  Such a large spatial separation of two distinct CO knots suggests that MIPS506 is a pair of merging galaxies.  The inferred dynamical masses are 5.4$\times$$\rm10^{11}$\,$\rm sin^{-2}(i)$ and 3.2$\times$$\rm10^{11}sin^{-2}(i)M_\odot$ for MIPS16144 and MIPS506 respectively, yielding the gas fraction ($\rm M_{gas}/M_{dyn}$) of 8\%\ and 6\%\ if using an averaged inclination angle. }

\item{Together with spatially resolved CO data, our HST/NICMOS H-band images suggest that majority of our sample have signatures of mergers at various stages of dynamical interactions, including close companions, tidal tails, and rings in the residual images.  We hypothesis that SMGs, bright 24\um\ ULIRGs and gas rich QSOs may be connected by the same evolutionary model in which the formation mechanism is primarily mergers, with SMGs mostly at an early passage of two gas rich, star forming galaxies, bright 24\um\ ULIRGs at a later stage of dynamical interaction, with gas and dust surrounding the central black holes, resulting smaller observed CO velocity width, and finally, the gas rich QSOs representing the final merged stage.  }

\end{itemize}


\acknowledgements  
Sylvain Veilleux is thanked for providing the electronic version of the mid-IR spectra for local {\it IRAS} 1\,Jy sample. D.  Riechers is gratefully acknowledged for very helpful discussions and for providing us a compiled list of  high-$z$ QSOs with  high resolution CO observations.  This work is based on observations with 30m telescope of the {\it Institute for Radioastronomy at Millimeter Wavelengths } (IRAM), which is funded by the German Max Planck Society, the French CNRS and the Spanish National Geographical Institute.  We thank the staff of the IRAM Observatory for their support of this program.  Also based on observations taken with {\it Spitzer Space Telescope}, which is funded by NASA and operated by  JPL/Caltech.  H-band imaging data are from observations made with the NASA/ESA Hubble Space Telescope and obtained at Space Telescope Science  Institute, operated by the Association of Universities for Research in Astronomy, Inc., under NASA contract NAS5-26555.  We thank helpful discussions with Nick Scoville and Arjun Dey.

\clearpage


\end{document}